\documentclass{article}

\usepackage{amsmath,amsfonts,bm}

\newcommand{\audiogenerator}{GenAu\xspace}
\newcommand{\audiocaptioner}{\mbox{AutoCap}\xspace}
\newcommand{\audiodataset}{\mbox{AutoReCap}\xspace}

\def\figref#1{Fig.~\ref{#1}}
\def\Figref#1{Figure~\ref{#1}}

\def\secref#1{section~\ref{#1}}

\def\eqref#1{equation~\ref{#1}}

\def\1{\bm{1}}

\DeclareMathAlphabet{\mathsfit}{\encodingdefault}{\sfdefault}{m}{sl}
\SetMathAlphabet{\mathsfit}{bold}{\encodingdefault}{\sfdefault}{bx}{n}

\newcommand{\ie}{\textit{i}.\textit{e}.}
\newcommand{\eg}{\textit{e}.\textit{g}.}

\newcommand{\name}{Ours}

\newcommand{\apref}[1]{Appx.~\ref*{#1}}
\newcommand{\tabref}[1]{Tab.~\ref{#1}}
\newcommand{\website}{\emph{Website}}
\newcommand{\supp}{\emph{Appendix}}

\newcommand{\qformer}{\mathcal{Q}}

\newcommand{\dectext}{\mathcal{D}_\mathrm{t}}
\newcommand{\encclap}{\mathcal{E}_\mathrm{clap}}
\newcommand{\projclap}{\mathcal{P}_\mathrm{clap}}
\newcommand{\encaudio}{\mathcal{E}_\mathrm{a}}

\newcommand{\textinput}{\mathbf{x}}
\newcommand{\audioinput}{\mathbf{x}_\mathrm{audio}}
\newcommand{\clapinput}{\mathbf{x}_\mathrm{clap}}

\newcommand{\metainputord}[1]{\mathbf{x}_{\mathrm{meta}_{#1}}}

\newcommand{\audio}{\mathbf{a}}

\newcommand{\metaord}[1]{\mathbf{m}_{#1}}

\newcommand{\textoutput}{\mathbf{\hat{y}}}
\newcommand{\textoutputgt}{\mathbf{y}}

\newcommand{\embeddingsflan}{e_{\mathrm{FLAN}}}
\newcommand{\embeddingsclap}{e_{\mathrm{CLAP}}}
\newcommand{\conditioning}{c}

\newcommand{\diffinput}{\mathbf{x}}

\newcommand{\nummeta}{M}

\newcommand{\tokenboa}{\texttt{\small[boa]}}
\newcommand{\tokeneoa}{\texttt{\small[eoa]}}

\newcommand{\tokenbom}[1]{\texttt{\small[bom]}_{#1}}
\newcommand{\tokeneom}[1]{\texttt{\small[bom]}_{#1}}

\newcommand{\difftimestep}{t}

\newcommand\nnfootnote[1]{%
  \begin{NoHyper}
  \renewcommand\thefootnote{*} %
  \footnotetext{#1}%
  \renewcommand\thefootnote{\arabic{footnote}} %
  \addtocounter{footnote}{-1}%
  \end{NoHyper}
}

\usepackage{xspace}
\usepackage{caption}
\usepackage{microtype}
\usepackage{graphicx}
\usepackage{subfigure}
\usepackage{booktabs} %

\usepackage{hyperref}

\usepackage[accepted]{icml2025}
\usepackage{amsmath}
\usepackage{amssymb}
\usepackage{mathtools}
\usepackage{amsthm}

\usepackage[capitalize,noabbrev]{cleveref}

\theoremstyle{plain}

\theoremstyle{definition}

\theoremstyle{remark}

\usepackage[textsize=tiny]{todonotes}

\begin{document}

\twocolumn[
\icmltitle{Taming Data and Transformers for Audio Generation}

\centerline{
    \textbf{Moayed Haji-Ali}$^{1,2,*}$ \quad
    \textbf{Willi Menapace}$^{2}$ \quad
    \textbf{Aliaksandr Siarohin}$^{2}$
}
\vspace{1em}
\centerline{
    \textbf{Guha Balakrishnan}$^{1}$ \quad
    \textbf{Vicente Ordonez}$^{1}$
}
\vspace{1em}

\centerline{${}^1\text{Rice University}$ \quad ${}^2\text{Snap Inc}$}
\vspace{1em}
\centerline{Project Webpage: \href{https://snap-research.github.io/GenAU}{\color{blue}https://snap-research.github.io/GenAU}}
\vskip 0.3in
]

\begin{abstract}
\nnfootnote{Work partially done during an internship at Snap Inc.}
The scalability of ambient sound generators is hindered by data scarcity, insufficient caption quality, and limited scalability in model architecture. This work addresses these challenges by advancing both data and model scaling. First, we propose an efficient and scalable dataset collection pipeline tailored for \emph{ambient} audio generation, resulting in AutoReCap-XL, the largest ambient audio-text dataset with over \emph{47 million} clips. To provide high-quality textual annotations, we propose \audiocaptioner, a \emph{high-quality} automatic audio captioning model. By adopting a Q-Former module and leveraging audio metadata, \audiocaptioner substantially enhances caption quality, reaching a CIDEr score of $83.2$, a $3.2\%$ improvement over previous captioning models. Finally, we propose \audiogenerator, a scalable transformer-based audio generation architecture that we scale up to 1.25B parameters. We demonstrate its benefits from data scaling with synthetic captions as well as model size scaling. When compared to baseline audio generators \emph{trained at similar size and data scale}, \audiogenerator obtains significant improvements of $4.7\%$ in FAD score, $11.1\%$ in IS, and $13.5\%$ in CLAP score. Our code, model checkpoints, and dataset are \emph{publicly available}.

\end{abstract}

\vspace{-1.5em}
\section{Introduction}
Text-conditioned generative models have revolutionized the field of content creation, enabling the generation of high-quality natural images~\citep{ramesh2022hierarchical,rombach2022high,podell2023sdxl, elasticdiffusion}, vivid videos~\cite{ho2022imagen,villegas2022phenaki,wang2023lavie,qiu2023freenoise,menapace2024snap}, 
and intricate 3D shapes~\citep{cheng2023sdfusion}.
The domain of audio synthesis has undergone comparable advancement~\citep{huang2023make,huang2023makeanaudio2,liu2023audioldm, auffusion, lafma, soundctm, niu2024soundlocd, uniaudio, evans2024fast, audiolcm, frieren, audiogeneration}, with three broad areas of study: speech, music and ambient sounds. 
The success in these domains rests on two key pillars: (i)~the availability of high-quality large-scale datasets with text annotations, and (ii)~the development of scalable generative models~\citep{ho2020denoising,song2021denoising}. The objective of this work is to improve audio generation quality by scaling \emph{ambient} audio generators across both the data and model axes.

In the field of audio synthesis, ambient audio generation emerges as a critical domain.
Unlike speech and music, ambient sound generation is particularly challenging due to the lack of extensive, well-annotated datasets~\citep{kim2019audiocaps,drossos2020clotho}. 
Attempts to curate ambient audio from online videos 
predominantly failed, primarily due to the dominance of speech and music content in such videos. For instance, AudioSet \citep{gemmeke2017audioset}, the largest available audio dataset sourced from online videos, contains $99\%$ speech or music clips. Previous efforts to filter ambient audio from similar datasets involved using expensive classifiers on the video or audio content, making it impractical to compile a large-scale dataset due to the high filtering rate. In this work, we propose a simple, yet scalable filtering approach that leverages existing automatic video transcription to identify segments with ambient sounds. This method is not only more efficient but also more feasible, as it eliminates the need to download audio or video content. Through this approach, we built \audiodataset-XL, a dataset containing 47 million ambient audio clips sourced from existing video datasets, representing a \emph{75-fold} increase over the size of previously largest available datasets.

Another challenge in compiling large-scale text-audio datasets is providing accurate textual descriptions. For visual modalities, such as images and videos~\citep{xue2022hdvila,miech19howto100m}, researchers often relied on the raw description and metadata to train strong visual-text models including reliable captioners~\citep{chen2024panda70m}. 
For ambient sounds, however, the task is substantially more challenging as accompanying raw text tends to describe visual information or convey feelings, rather than detailing the audio content. Moreover, human-captioned audio datasets are limited, containing fewer than $51k$ text-audio pairs in total. This significantly impacts the training of current captioning models, making them more susceptible to overfitting. To address this, we introduce \audiocaptioner, a high-quality audio captioner that leverages visual cues to enhance captioning.

\audiocaptioner refines the commonly used encoder-decoder design based on a pretrained BART~\citep{lewis2020bart} by introducing a Q-Former~\citep{li2023blip2} module which learns an intermediate representation that better aligns the encoded audio and the original BART token representation. Second, we propose to remedy the data scarcity problem by using metadata and visual cues to aid the captioning process. Critically, we augment the encoder inputs with a set of descriptive textual metadata including audio title and a caption derived from the visual modality. This dual-input approach achieves state-of-the-art performance on AudioCaps \citep{kim2019audiocaps}, marking a 3.2\% improvement in CIDEr score. Using \audiocaptioner, we provide textual descriptions for \audiodataset-XL and demonstrate the benefits of scaling audio generative models with synthetic captions.

Another axis for scaling generative models is model size~\cite{dit}. While scaling diffusion backbones has shown significant benefits in image and video generation, ambient audio generation has shown poor scaling behavior. For instance, AudioLDM2~\cite{liu2023audioldm2}, reported worse metrics for their largest model compared to the smaller one. Similarly, EzAudio~\cite{ezaudio}, achieved only marginal improvements with scaling model size. In this work, we introduce \audiogenerator, a \emph{scalable} transformer-based architecture that achieves \emph{significant} improvements over state-of-the-art models. We recognize that audio grows fast temporally, yet contains many silent and redundant segments. Therefore, an efficient architecture that can handle such properties is needed. In particular, we employ a transformer architecture in the denoising backbone where we modify the FIT transformer \citep{chen2023fit} to generate audio in the latent space.
On AutoCaps dataset, \audiogenerator achieves significant improvements over baselines, with \emph{$11.1\%$} higher Inception Score, \emph{$4.7\%$} better FAD, and \emph{$13.5\%$} improvement in CLAP score, demonstrating superior audio-text alignment and generation quality. Moreover, \audiogenerator shows promising scaling properties, with consistent improvements across all metrics as model size increases.

In summary, this work presents significant contributions in three areas: (i) \audiocaptioner, a novel \emph{state-of-the-art} audio captioner tailored towards the annotation of data at a large scale that uses audio metadata to improve accuracy and robustness;  (ii) \audiodataset-XL, a large scale ambient audio dataset, comprising 47M audio clips paired with synthetic captions, 75 times larger than available datasets (iii) \audiogenerator, a novel audio generator based on a scalable transformer architecture specifically adapted to the audio domain, achieving significant improvements over previous state-of-the-art.

\begin{figure*}
    \centering
    \includegraphics [width=\linewidth]{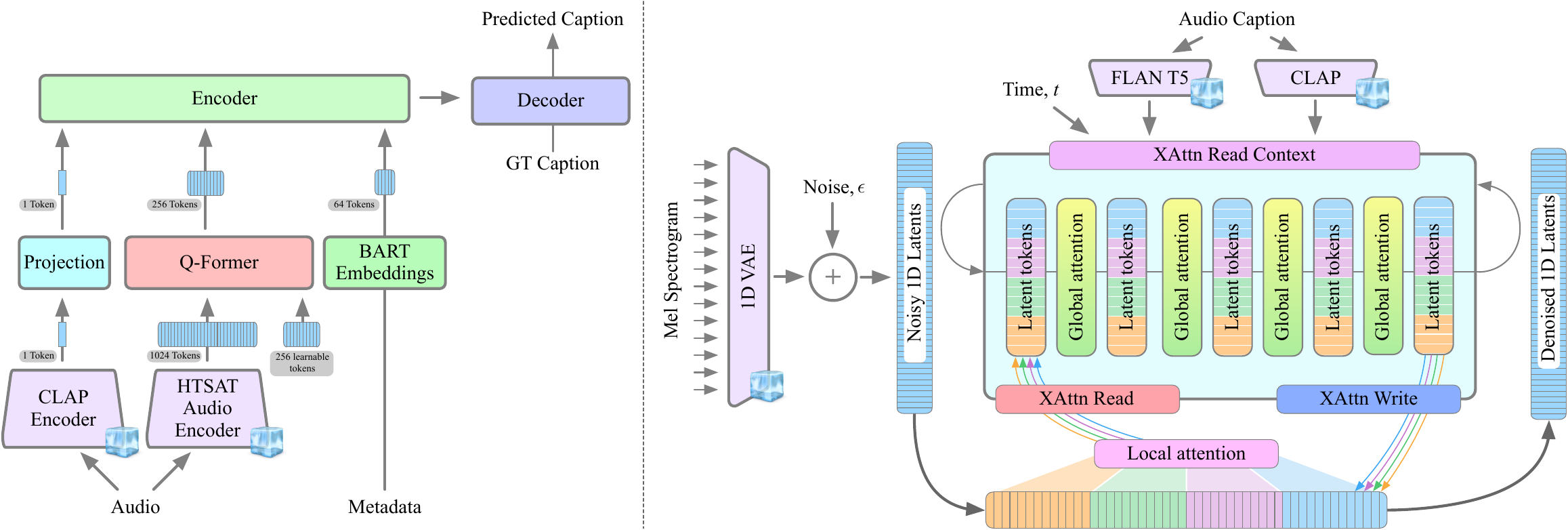}
    \caption{\textbf{(Left) Overview  of \audiocaptioner.} 
    We employ a frozen HTSAT~\citep{ke2022htsat} encoder to produce an audio fine-grained representation of 1024 tokens. We then employ a Q-Former~\citep{li2023blip2} module to produce 256 tokens. These tokens, along with audio CLAP embeddings~\citep{clap_laion} and 64 tokens of pertinent metadata, are processed by a pretrained BART to generate the final caption. \textbf{(Right) Overview of \audiogenerator.} Following latent diffusion models, we use a frozen 1D-VAE to convert a Mel-Spectrogram into latent sequences, which are then divided into groups and processed using `local' attention layers based on the FIT architecture~\citep{chen2023fit}. `Read' and `write' layers, implemented as cross-attention, facilitate information transfer between input latents and \emph{learnable} latent tokens. Finally, `global' attention layers on \emph{latent tokens} facilitate global communication across groups.}
    \label{fig:framework}
\end{figure*}
\section{Related Work}

\textbf{Automatic Audio Captioning (AAC).} The goal of AAC is to produce language descriptions for given audio content. 
Recent AAC methods~\citep{trainingaudiocaptioningmodels, gontier2021automated, wu2024improving, salewski2023zeroshot, sridhar2023parameter, kadlčík2023whisper, multilingual, labbé2023killing, xu2023secap, zhang2024zeroshot, recap_retrievalaugmentedaudiocaptioning, trainingaudiocaptioningmodels} employ encoder-decoder transformer architectures, where an encoder receiving the audio signal produces a representation that is used by the decoder to produce the output caption. WavCaps~\citep{mei2023wavcaps} employs HTSAT~\citep{ke2022htsat} audio encoder and uses a pretrained BART~\citep{lewis2020bart} as the decoder. Similarly, EnCLAP~\citep{kim2024enclap} uses a pretrained BART and improves on the audio representation. Recognizing the limited data, CoNeTTE~\citep{labbe2023conette} proposes to train a lightweight vanilla transformer decoder~\citep{vaswani2017attention} instead. Other work explored augmentation to counter data scarcity~\citep{kim2022exploring,labbe2023conette,ye2022featurecut}. Recent work~\citep{liu2024interspeech, autoacd, soundvecapsimprovingaudiogeneration} proposed to leverage visual information to address sound ambiguities, reporting improvements. More recent methods use audio large language model for zero-shot captioning~\cite{audioflamingonovelaudio, deshmukh2024pengiaudiolanguagemodel, gama_audio}. Our method uses audio metadata and visual information as additional signals and leverages a \emph{lightweight} Q-Former~\citep{li2023blip2} model to improve accuracy.

\textbf{Text-conditioned audio generation.} The current state-of-the-art text-to-audio generation methods widely adopt diffusion models~\citep{yang2023diffsound,kreuk2023audiogen,liu2023audioldm,liu2023audioldm2,huang2023makeanaudio2,ghosal2023text,evans2024fast,vyas2023audiobox, kreuk2023audiogen, ezaudio, stableaudioopen}. 
AudioLDM 1 \& 2~\citep{liu2023audioldm, liu2023audioldm2} make use of a latent diffusion model and employ a UNet as the diffusion backbone.
Recently, StableAudio Open~\citep{stableaudioopen} introduced a 1.32B model that uses a DiT~\citep{dit} to generate variable-length audio clips at 48kHz.
Recent work also explored controllable audio generation~\citep{shi2023enhance, xu2024promptguided, melechovsky2024mustango, paissan2024audio, zhang2023loop, liang2024wavcraft, liu2023audiosr}, visual-conditioned audio generation~\citep{frieren, mei2023foleygen, wang2023v2amapper}, and more recently joint audio-video generation~\citep{tang2023codi2, tang2023anytoany, xing2024seeing, hayakawa2024discriminatorguided, tian2024vidmuse, vahdati2024deepfake, chen2024semantically, kim2024versatile, wang2024avdit, mao2024tavgbench, avlink}. In this work, we propose a transformer architecture design that shows strong scalability properties.

\textbf{Text-Audio Datasets.}
The performance of text-audio models~\citep{zhu2024cacophony, li2023audiofree, deshmukh2024pengiaudiolanguagemodel, mahfuz2023improving, deshmukh2024domain, shu2023audiovisual, elizalde2024natural, liu2023separate, tang2024salmonn, gong2024listenthinkunderstand, cheng2024videollama, zhang2023videollama} is currently hindered by the lack of high-quality large-scale paired audio text data of ambient sounds. Existing human-captioned (AudioCaps~\citep{kim2019audiocaps} and Clotho~\citep{drossos2020clotho}), have in total only $51k$ audio-text pairs. Another challenge is the limited availability of audio clips from sound-only platforms. LAION-Audio~\citep{clap_laion} relied on numerous sources of audio platforms such as BBC Sound Effects~\citep{bbc_sound_effects}, ~\citep{freesounds} FreeSounds, and SoundBible~\citep{soundbible} to form a dataset consisting of 630k audio samples with highly noisy raw descriptions. \citet{chen2020vggsound} attempted to extract audio clips from videos by employing classifiers to detect ambient audio, speech, and music. To annotate these clips, WavCaps~\citep{mei2023wavcaps} proposes a filtering procedure based on ChatGPT~\citep{openai2024gpt4} to collect $400k$ audio clips and weakly caption them \emph{based on the noisy descriptions alone.} Several subsequent work \citep{tango2, autoacd} adopted similar strategies of using large language models to augment captions. While weak-captioning improves downstream metrics, it is \emph{suboptimal} because it fails to incorporate the audio signal itself. 
In this work, we introduce an efficient dataset collection pipeline that relies on video datasets to extract ambient audio clips and automatic captioners to provide textual descriptions. We collect $47M$ audio clips, marking the largest available text-audio dataset.

\begin{figure*}
    \centering
    \includegraphics[width=\linewidth]{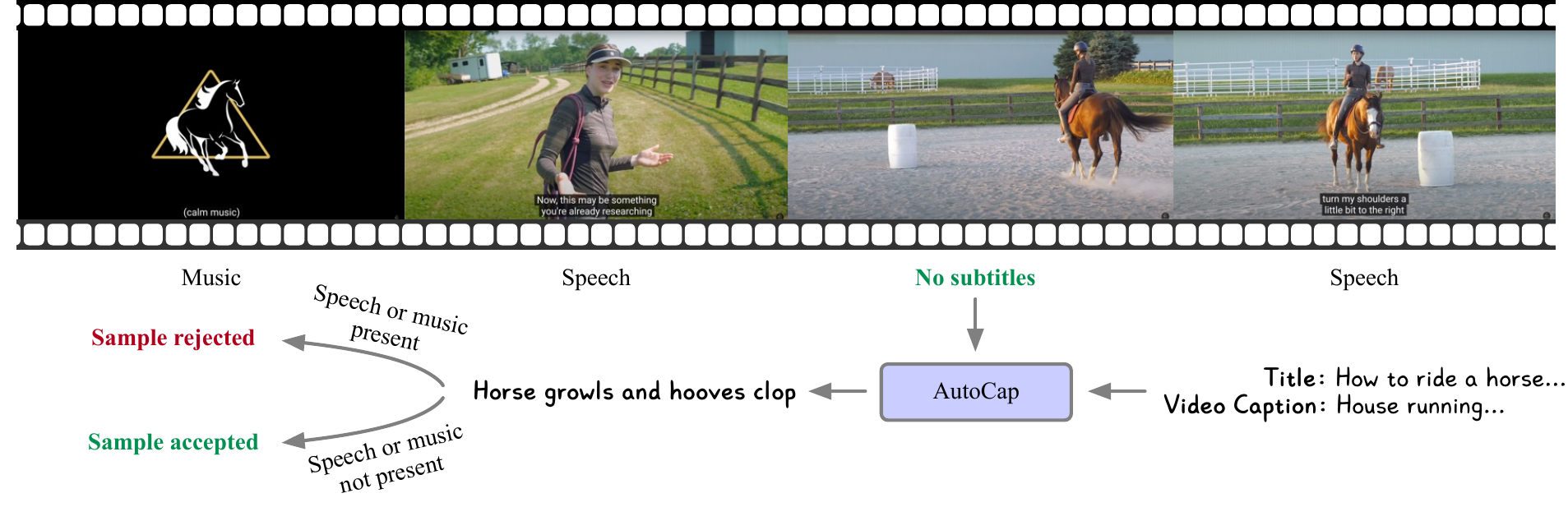}
    \caption{\textbf{Audio data collection pipeline.} We employ online video transcripts to identify audio segments without speech or music. These are processed by \audiocaptioner to generate captions. We retain only ambient clips with captions lacking music and speech keywords.}
    \label{fig:data_pipeline}
\end{figure*}

\section{Method}
\label{sec:method}
In this section, we describe our approach to high-quality text-to-audio generation, starting with audio captioning using~\audiocaptioner in \secref{sec:automatic_audio_captioning}, data collection in \secref{sec:audio_recaptioning}, and ambient audio generation with \audiogenerator in~\secref{sec:audio_generation}  
\subsection{Automatic Audio Captioning}
\label{sec:automatic_audio_captioning}

Recent state-of-the-art methods \citep{labbe2023conette, kim2024enclap} generally employ an encoder-decoder transformer design where a pretrained audio encoder passes the audio representation to a pre-trained language model serving as the decoder. This language model (\eg~BART) is typically finetuned to adapt to the audio representation. However, due to the distribution mismatch between the pretraining data of the LLM and the audio embeddings produced by the encoder, the decoder suffers from catastrophic forgetting. Furthermore, audio is an inherently ambiguous modality, as many events can produce similar sound effects—a phenomenon often leveraged in animation, where soundscapes are artificially constructed. Audio clips from many sources, however, are still commonly associated with metadata that might be relevant for captioning such as raw user descriptions, or related modalities ({\em i.e.}~accompanied visual information). Motivated by these observations, we propose \audiocaptioner, an audio captioning model that employs an intermediate audio representation to connect the pretrained encoder and decoder and uses metadata to aid with the captioning. \Figref{fig:framework} (left) presents an overview of \audiocaptioner. 

We consider a dataset of audio-caption pairs $\langle\audio, \textoutputgt\rangle$ and corresponding metadata represented as a set of token sequences $\{\metaord{j}\}_{j=1}^{j=\nummeta}$.
Inspired by state-of-the-art AAC methods~\citep{mei2023wavcaps,labbe2023conette,kim2024enclap}, we employ an encoder-decoder architecture. We first compute a global feature representation of the audio: 
\begin{equation}
\clapinput=\projclap(\encclap(\audio)),
\end{equation}
where $\projclap$ is a learnable projection layer and $\encclap$ is the audio encoder of a pretrained CLAP model~\citep{clap_laion}. We also compute local features as: 
\begin{equation}
\audioinput=\qformer(\encaudio(\audio)),
\end{equation}
where $\qformer$ is a Q-Former~\citep{li2023blip2} and $\encaudio$ is a pretrained HTSAT~\citep{ke2022htsat} audio encoder that produces a \emph{time-aligned representation} (1024 tokens) following~\cite{mei2023wavcaps}. The Q-Former efficiently learns 256 latent tokens, which serve as keys in cross-attention layers with the input features, thereby outputting 256 tokens. Metadata sequences $\metaord{i}$ are then embedded using the embedding layer of a pretrained BART to obtain embedding sequences $\metainputord{i}$. For our experiments, we use video titles and captions as the metadata. We represent the input audio and metadata as the following input sequence: 
\begin{equation}
    \begin{aligned}
        \textinput &= \clapinput~\tokenboa~\audioinput~\tokeneoa ~\tokenbom{1}~\metainputord{1} \\
        &\quad~\tokeneom{1}~...~\tokenbom{\nummeta}~\metainputord{\nummeta}~\tokeneom{\nummeta}
    \end{aligned}
    \text{,}
\end{equation}
where $\tokenboa\tokeneoa$ represent beginning and end of audio sequence embeddings $\audioinput$, and $\tokenbom{i}, \tokeneom{i}$ represent beginning and end of metadata embeddings $\metainputord{i}$. The input sequence is passed to a 
pretrained BART~\citep{lewis2020bart} $\dectext$ to predict a caption as $\textoutput = \dectext(\textinput)$

\textbf{Training.} We train our model using a standard cross-entropy loss over next token predictions. To avoid degrading the quality of the pretrained BART and audio encoder models, we adopt a two-stage training procedure. In Stage 1, both the audio encoders and BART model are kept frozen, thus allowing the Q-Former, projection layers, and newly introduced delimiter tokens to align to the pretraining BART representation. In this stage, we pretrain the model using a larger dataset of weakly-labeled audio clips. In Stage 2, we unfreeze all BART parameters apart from the embedding layer and finetune the model on the Audiocaps dataset at a lower learning rate to align the captioning style more with human style. This training strategy leverages the larger, weakly-labeled dataset while minimizing the knowledge drift in the pretrained BART. The use of Q-Former to learn an intermediate representation is \emph{pivotal} for such a strategy.

\begin{figure*}
    \centering
    \includegraphics[width=\linewidth]{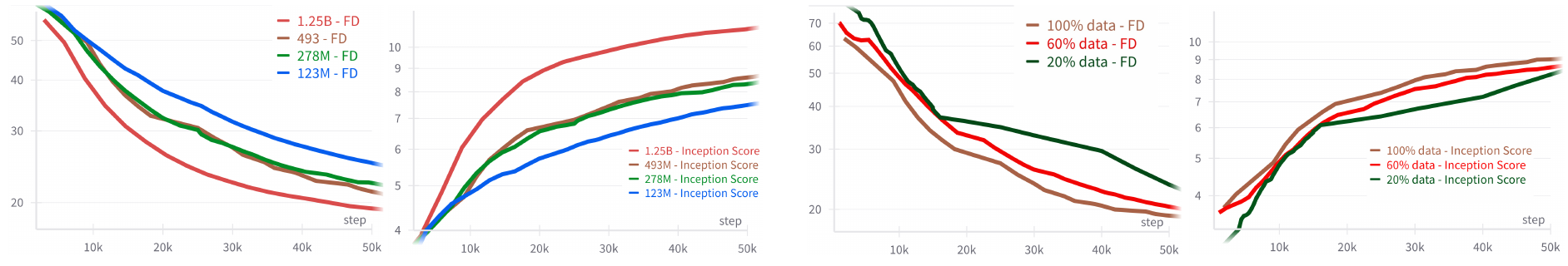}
    \caption{\textbf{Scaling analysis} of model size (left) and data with synthetic captions (right) reveal consistent improvements in FD and IS.}
    \label{fig:scaling}
    \vspace{-1em}
\end{figure*}

\subsection{Data Collection and Re-captioning Pipeline}
\label{sec:audio_recaptioning}
Generative models in the image and video domains have shown benefits from increased quantities of data and improved quality of captions. In the audio domain, however, the major human-annotated audio-text datasets, namely AudioCaps \citep{kim2019audiocaps} and Clotho \citep{drossos2020clotho}, provide only \emph{51k} audio clips combined.  Previous methods attempted to extract additional ambient audio clips from existing video datasets using pretrained audio classifiers, but a high filtering rate marked this method impractical. Instead, we found that automatic transcripts offer reliable information about the segments containing ambient sounds. In particular, we propose to select the parts of the videos that contain no automatic transcription, suggesting the absence of speech and music. Such an approach offers specific advantages over using pretrained classifiers. Automatic transcripts, readily available for most online videos, eliminate the need to download and process video and audio data before filtering. Additionally, as these transcripts provide precise time-aligned information, they facilitate the extraction of more segments per video. Subsequently, we leverage our \audiocaptioner model to provide textual descriptions of the extracted audio clips. Despite the effectiveness of this method in collecting ambient sounds, some clips still inadvertently contain music or speech due to transcription errors, particularly with speech in less common languages. We address this by analyzing captions and filtering out clips with keywords related to speech or music. Finally, we filter all audio-text pairs with CLAP similarity less than 0.1. 

We follow this process to extract 466k audio-text pairs from Audioset \citep{gemmeke2017audioset} and VGGSounds \citep{chen2020vggsound}. Additionally, we recaption audio-only datasets such as  Freesound, BBC Sound Effects, and SoundBible. To provide metadata, we employ the captioning model of  \citet{chen2024panda70m} to extract a caption whenever video content is available and pass an empty text otherwise. In total, we form \audiodataset, a large-scale dataset compromising of \emph{761,113} audio-text pairs with precise captions. As an additional contribution, we introduce AutoReCap-XL, in which we scale our approach by analyzing four additional large-scale video dataset \citep{lee2021acav100mautomaticcurationlargescale, xue2022hdvila, ytyemp1b, nagrani2022learningaudiovideomodalitiesimage} with a total of \emph{71M} videos and \emph{$715.4k$} hours. After filtering, we collect and re-caption \emph{47M} ambient audio clips spanning \emph{$123.5k$} hours from \emph{20.3M} different videos, forming by far the largest available dataset of audio with paired captions. \Figref{fig:data_pipeline} summarizes our data collection pipeline, and Sec.~\ref{sec:autorecap-xl} in \supp~present more details about the dataset collection and processing.

\begin{table*}[t!]
\centering
\small
\setlength\tabcolsep{1.8pt}
\renewcommand{\arraystretch}{1.1}
\caption{\audiocaptioner results on AudioCaps test split for various models. AS: AudioSet, AC: AudioCaps, WC: WavCaps, CL: Clotho, MA: Multi-Annotator Captioned Soundscapes.}
\label{tab:captioning-main}
\resizebox{0.8\textwidth}{!}{%
\begin{tabular}{ll c ccccccc}
\toprule
 {Model} & {Pretraining Data} && BLEU1 & BLEU4 & ROUGE$_L$ & METEOR & CIDEr & SPICE & SPIDEr \\ 
\midrule %
V-ACT & -          && 69.8  & 28.1  & 49.4      & 23.7   & 71.1  & 17.2  & 44.2   \\
BART-tags & AS     && 69.9  & 26.6  & 49.3      & 24.1   & 75.3  & 17.6  & 46.5   \\
AL-MixGEN & -      && 70.0  & 28.9  & 50.2      & 24.2   & 76.9  & 18.1  & 47.5   \\
GAMA & - && 20.4 & 12.2 & 19.1 & 16.3 & 64.8  & 5.4 & 35.1	\\
ENCLAP-Large & -        && -  & - & -      & 25.5  & 80.2 & 18.8 & 49.5   \\ 
HTSAT-BART  & -      && 67.5  & 27.2  & 48.3      & 23.7   & 72.1  & 16.9  & 44.5   \\
HTSAT-BART & AC+CL+WC    && 70.7  & 28.3  & 50.7      & 25.0   & 78.7  & 18.2  & 48.5   \\
CNext-trans & -    && - & -     & -  & - & - & - &  46.6  \\
CNext-trans & AC+CL+MA+WC        && -  & - & -     & 25.2  & 80.6 & 18.4  & 49.5   \\

\midrule

\audiocaptioner~(audio) & AC       && 70.0 & 28.0  & 51.7      & 24.6  & 77.3  & 18.2   & 47.8  \\
\audiocaptioner~(audio+text) & AC  && 72.1  & 28.6 & 51.5      & \textbf{25.6}   & 80.0  & 18.8   & 49.4  \\
\audiocaptioner~(audio) & AC+CL+WC   && \textbf{73.1} & 28.1  & \textbf{52.0}  &  \textbf{25.6} & 80.4  & \textbf{19.0}   & 49.7 \\
\audiocaptioner~(audio+text) & AC+CL+WC  && 72.3 & \textbf{29.7}  & 51.8   & 25.3  & \textbf{83.2}  & 18.2   & \textbf{50.7}  \\

\bottomrule
\end{tabular}%
}
\end{table*}

\subsection{Scalable Text-2-Audio Generation}
\label{sec:audio_generation}

We design our audio generation pipeline, \audiogenerator, as a latent diffusion model. \Figref{fig:framework} (right) shows an overview of our proposed model. %
In the following section, we describe in detail the structure of our latent variational autoencoder (VAE) and the latent diffusion model.

\noindent\textbf{Latent VAE. } 
Directly modeling waveforms is complex due to the high data dimensionality of audio signals. Instead, we replace the waveform with a Mel-spectrogram representation and use a VAE to further reduce its dimensionality, following prior work~\citep{melechovsky2024mustango, huang2023make}. %
Once generated, Mel-spectrograms can be decoded back to a waveform through a vocoder~\citep{bigvgan}. However, commonly-used 2D autoencoder designs \citep{liu2023audioldm, liu2023audioldm2, melechovsky2024mustango}, are not well suited to the Mel-spectrograms, as the separation between the Mel channels is non-linear, which is not well suited for 2D convolutions. We instead opt for a 1D-VAE design based on 1D convolutions similar to~\citet{huang2023makeanaudio2}. We train the VAE following~\citet{esser2021taming}.

\noindent\textbf{Latent diffusion model.}
Following the latent diffusion paradigm, we generate audio by training a diffusion model in the latent space of the 1D-VAE. Transformer-based diffusion models currently attain state-of-the-art performance in audio generation~\citep{huang2023makeanaudio2}. However, both UNet and transformer-based baselines exhibited limited performance gains with increasing model size~\cite{liu2023audioldm2, ezaudio}. We observe that ambient audio often contains extensive silent and redundant segments, which may explain the poor scalability of UNet and DiT-based methods, as they distribute computation uniformly across the input. Therefore, we propose to use a more dynamic transformer architecture as a diffusion backbone~\citep{chen2023fit,menapace2024snap}. In particular, we adopt the FIT architecture of~\citet{menapace2024snap} which was originally proposed to work in the \emph{pixel space}, and revise it for the \emph{latent space} of the audio modality.

Given a 1D input $\diffinput$, we first apply a projection operation to produce a sequence of input patch tokens. We then apply a sequence of FIT blocks to the input patches where each block divides patch tokens into contiguous groups of a predefined size. A set of \emph{local} self-attention layers are then applied separately to each group to avoid the quadratic computational complexity of attention computation. Differently from the video domain~\citep{menapace2024snap} where the high input dimensionality makes the \emph{local} layers excessively expensive, we found them to be beneficial for audio generation. To further reduce the amount of computation while maintaining long-range interaction, each block considers a small set of latent tokens. First, a \emph{read} operation implemented as a cross-attention layer transfers information from the patches to the latent tokens. Later, a series of \emph{global} self-attention operations are applied to the latent tokens, allowing information-sharing between different groups. Finally, a \emph{write} operation implemented as a cross-attention layer transfers information from the latent tokens back to the patches. Due to the reduced number of latent tokens when performing the global self-attention, computational requirements of the model are reduced with respect to a vanilla transformer design~\citep{vaswani2017attention}. Such a design is particularly suited for the audio modality, which contains mostly silent or redundant parts. Unlike DiT and UNet-based methods \citep{ronneberger2015unet, dit} which allocate the computation resources uniformly across input tokens, the FTT architecture selectively focuses on the more informative parts,  dedicating more compute for these parts as the model size scales. 

To condition the generation on an input prompt, we use a pretrained FLAN-T5 model~\citep{chung2022scaling} and a CLAP~\citep{clap_laion} text encoder to produce the their respective embeddings $\embeddingsflan$ and $\embeddingsclap$ following prior work \cite{liu2023audioldm2}, which we concatenate with the diffusion timestep $\difftimestep$ to form the input conditioning signal~$\conditioning$. We insert an additional cross-attention operation inside each FIT block immediately before the `read' operation that makes latent tokens attend to the conditioning. Moreover, we use conditioning on dataset ID to adapt the generation style to different datasets. We train the model using the epsilon prediction objective and follow a linear noise scheduler.

\section{Experiments}
\label{sec:experiments}
\begin{table}[]
    \centering
    \caption{\audiocaptioner ablation study on AudioCaps}
    \resizebox{\linewidth}{!}{
    \begin{tabular}{@{}lcccc@{}}
\toprule
Model                & METEOR $\uparrow$ & CIDEr $\uparrow$ & SPICE $\uparrow$ & SPIDEr $\uparrow$ \\ \midrule
\name              & \textbf{25.3}  & \textbf{83.2}  & 18.2   & \textbf{50.7}        \\
- w/o CLAP               & \textbf{25.3}               & 80.7             & \textbf{18.4}             & 49.6                \\
- w/o Stage 2           & 24.2             & 75.6             & 17.3             & 46.5               \\
- w/o Stage 1             & 22.6              & 59.6             & 15.4             & 37.5               \\
- Unfreeze Word Embed                 & 22.5               & 82.6                &   18.1         & 50.4                \\
\bottomrule
\end{tabular}}
    \label{tab:captioning-ablation}

\vspace{-1em}
\end{table}

In~\secref{sec:experiments_aac}, we evaluate \audiocaptioner quantitatively. We then demonstrate the capabilities of \audiogenerator in~\secref{sec:experiments_audio_generation} and discuss scaling trends with respect to data and model size. We also provide qualitative comparisons on the \website. 

\subsection{Automatic Audio Captioning}
\label{sec:experiments_aac}
\textbf{Training dataset and details.} We train \audiocaptioner in two stages. During stage 1, we pretrain on a large weakly labeled dataset of 634,208 audio clips, constructed from AudioSet, Freesound, BBC Sound Effects, SoundBible, AudioCaps, and Clotho. We use ground truth captions from AudioCaps and Clotho datasets, WavCaps captions for Freesound, SoundBible, and BBC Sound Effects, and handcrafted captions through a template leveraging the ground truth class labels for AudioSet. As metadata, we use the title provided with each clip, and pre-compute video captions using a pretrained Panda70M model~\citep{chen2024panda70m} or pass an empty string when the video modality is unavailable. We pretrain the model for 20 epochs with a learning rate of 1e-4, while keeping the audio encoder and pretrained BART frozen. In Stage 2, we fine-tune the model for 20 epochs on AudioCaps using a learning rate of 1e-5. We use 10-second clips at 32KHz for all experiments.

\textbf{Baselines.} We compare with V-ACT~\citep{liu2024interspeech}, BART-tags~\citep{gontier2021automated}, AL-MixGEN~\citep{kim2022exploring}, ENCLAP~\citep{kim2024enclap}, HTSAT-BART~\citep{xu2023secap}, CNext-trans~\citep{labbe2023conette} and GAMA~\cite{gama_audio}. %
Among these, ENCLAP and CNext-trans have the best performance. ENCLAP benefits from a stronger audio encoder and a CLAP representation. CNext-trans trains a lightweight transformer instead of fine-tuning a pretrained language model to reduce overfitting. 

\textbf{Metrics and evaluation.} We report results using the established BLEU1 and  BLEU4~\citep{papineni2002bleu}, ROUGE~\citep{lin2004rouge}, Meteor~\citep{lavia2007meteor}, CIDEr~\citep{vedantam2015cider}, and SPIDEr~\citep{liu2017spider} metrics. We evaluate our method on the AudioCaps test split using the last checkpoint of our trained model. 
We follow the same evaluation pipeline as baselines and include their reported results, except for GAMA which we evaluate using their released checkpoint. Metrics unavailable in these publications are excluded from our analysis.

\begin{table*}[t]
\setlength\tabcolsep{4pt}
\renewcommand{\arraystretch}{1.1}
\small
\centering
\caption{\audiogenerator results on AudioCaps test split.}
\label{tab:audio_generation_results}
\resizebox{0.8\textwidth}{!}{%
\begin{tabular}{@{}llcccccccc@{}}
\toprule
Model            & Prams  & $\#$ Samples & FD $\downarrow$  & IS $\uparrow$  & FAD $\downarrow$ & $\text{CLAP}_{LAION}$ $\uparrow$ & $\text{CLAP}_{MS}\uparrow$  \\ \midrule
GroundTruth      & -  & -     & -   & -   & -    &  0.251 & 0.671 \\
AudioLDM-L  & 739M   & 634k & 37.89 & 7.14 &  5.86 & - & 0.429 \\
AudioLDM 2-L  & 712M  & 760k & 32.50 & 8.54 &  5.11 & 0.212 & 0.621  \\
TANGO           &  866M & 45k & 26.13 & 8.23 &  1.87 & 0.185
& 0.597 \\
TANGO 2  & 866M & 60k & 19.77 & 8.45 & 2.74 & 0.264 & 0.590 \\
Make-An-Audio  & 453M    & 1M & 27.93 & 7.44 &  2.59 & 0.207 & 0.621  \\
Make-An-Audio 2  & 937M  & 1M  & 15.34 & 9.58 & 1.27 & 0.251 & 0.645 \\
Stable Audio Open  & 1.32B  & 486K  & 21.23 & 10.48 & 2.32 & 0.246 & 0.584 \\
\midrule
\audiogenerator-Large  & 1.25B & 811K & 16.51  & 11.75 & \textbf{1.21} & 0.285 & 0.668 \\
\audiogenerator-Large-Full  & 1.25B & 19.6M & \textbf{14.48}  & \textbf{12.56} & 1.71 &\textbf{0.318} & \textbf{0.677} \\
\bottomrule
\end{tabular}
}
\end{table*}

\textbf{Results.} Tab.~\ref{tab:captioning-main} reports that our method outperforms baselines on all metrics, achieving notable improvements in CIDEr (83.2) and BLUE1 (73.1) scores. Notably, even without metadata (\ie~\emph{audio only}), \audiocaptioner surpasses baselines in most metrics.
We found that incorporating metadata significantly enhances CIDEr but slightly reduces SPICE. This trade-off likely results from the enhanced descriptive detail brought by the metadata, which while enriching the content, introduces noise that may compromise the model's semantic precision. In addition, AudioCaps is labeled based on audio alone. Thus, the evaluation penalizes the description of information that can not be deduced with certainty from the audio modality only, such as the specific type of object producing the sound. Qualitatively, our captions are more detailed and temporally accurate than baselines. ENCLAP-Large often misses key details. CNext-trans, while accurate, often produces short captions that lack details. We include qualitative comparisons in the~\website~and~\supp. Moreover, \audiocaptioner is \emph{four times} faster than ENCALP, producing a caption for a 10-second clip in \emph{0.28} seconds, compared to ENCALP which takes \emph{1.12} seconds. Furthermore, we observe consistent improvements when pretraining on weakly-labeled data, validating the effectiveness of our training strategy in benefiting from larger, weakly-labeled datasets.

\textbf{Ablations.} 
In \tabref{tab:captioning-ablation}, we ablate model design choices. Using CLAP embedding brings a $2.5$ points increase in CIDEr. Omitting Stage 2 training, which involves finetuning  BART~\citep{lewis2020bart}, results in performance degradation, likely due to the necessity of adapting BART's decoder to the sentence structure typical of AudioCaps. A more severe degradation in performance is observed when Stage 1 is not performed, as the misaligned representation between the encoder and the decoder leads to catastrophic forgetting in the language model. 
Finally, finetunning BART word embeddings in Stage 2 reduces performance slightly.
\subsection{Text-2-Audio Generation}
\label{sec:experiments_audio_generation}

\textbf{Training dataset and details.} We follow baselines~\cite{liu2023audioldm2, huang2023makeanaudio2, tango2} and train on 10-second clips at 16kHz resolution. We use a patch size of 1 and a group size of 32. We use LAMB optimizer~\citep{lamb} with a LR of 5e-3. We train for 220k steps and choose the checkpoint with the highest IS.

\textbf{Baselines.} We compare with TANGO 1 \& 2,~\citep{ghosal2023text}, AudioLDM 1 \& 2~\citep{liu2023audioldm,liu2023audioldm2}, and Make-An-Audio 1 \& 2~\citep{huang2023make,huang2023makeanaudio2}. Both AudioLDM and Make-an-Audio train a UNet-based latent diffusion model~\citep{rombach2022high} on Mel-Spectrogram representation, by regarding it as a single channel image, and use a pretrained CLAP encoder to condition the generation on an input prompt. TANGO proposed to use FLAN-T5~\citep{chung2022scaling} as the text encoder and reported significant improvements. AudioLDM-2 and Make-an-Audio-2 proposed to use a dual encoder strategy of a T5~\citep{raffel2022exploring} and CLAP encoder. 
Make-an-Audio-2 proposes to use a 1D VAE representation and employ a DiT as the diffusion backbone.
Recently, Tango-2 proposed to use instruction fine-tuning on a synthetic dataset to enhance temporal understanding. In our experiments, we focus on text-conditioned natural audio generation.

\textbf{Metrics.} We compare the performance of our method with baselines using the standard Frechet Distance (FD), Inception score (IS), and CLAP score on the AudioSet test split. There is little consistency between baselines when computing the metrics. Some prior work reported the Fréchet distance results using the VGGish network~\citep{hershey2017vggish}, denoted as (FAD)~\citep{kilgour2019frechet}, while other uses PANNs~\citep{kong2019pann}. Additionally, to compute the CLAP score, some prior work~\citep{liu2023audioldm2} used CLAP from LAION, which we denote as $\text{CLAP}_{LAION}$~\citep{wu2023laionaudio}, while others~\citep{tango2, huang2023make, huang2023makeanaudio2} used CLAP from Microsoft~\citep{elizalde2023clap}, which we denote as $\text{CLAP}_{MS}$. Furthermore, some prior~\citep{liu2023audioldm, liu2023audioldm2} used CLAP re-ranking with 3 samples for computing the metrics. Due to such inconsistencies in evaluation pipelines and varying results for the same baselines reported in different studies, we recompute all metrics using the official checkpoints to ensure consistent comparisons. We follow the same evaluation protocols of AudioLDM~\citep{liu2023audioldm} without CLAP re-ranking and use the AudioLDM evaluation package to compute the metrics. Besides, to prevent biasing the evaluation based on the training data, we run our ablations on the Bigsoundbank split from WavText5k~\citep{wavtext5k}, which serves as an out-of-distribution evaluation for our models. Finally, to further validate our results, we run user studies. Details about the user study can be found in the \supp.  
\begin{table}[t]
\centering
\caption{\audiogenerator ablation study on out-of-distribution dataset.}
\label{tab:ood_evaluation}
\resizebox{\linewidth}{!}{
\begin{tabular}{@{}llccc@{}}
        \toprule
        Ablation & Model & IS & FD & $\text{CLAP}_{MS}$ \\
        \midrule
        Model Scale &&&&
        \vspace{-0.75em}\\
        &GenAU-S & 15.76 & 21.29 & 0.36 \\
        &GenAU-L & \textbf{18.98} & \textbf{20.81} & \textbf{0.38} \\
        \midrule
        Data Scale &&&& 
        \vspace{-0.75em}\\
        &GenAU-L (AC) & 12.14 & 25.82 & 0.30 \\
        &GenAU-L (AutoReCap) & \textbf{18.98} & \textbf{20.81} & \textbf{0.38} \\
        \midrule
        Synthetic Captions &&&&
        \vspace{-0.75em}\\
        &GenAU-S w/o Recap. & 11.83 & 25.34 & 0.29 \\
        &GenAU-S & \textbf{15.76} & \textbf{21.29} & \textbf{0.36} \\
        \midrule
        Backbone &&&&
        \vspace{-0.75em}\\
        &GenAU-S w/ U-Net & 13.55 & 32.51 & 0.28 \\
        &GenAU-S w/ DiT & 13.96 & 23.32 & 0.33 \\
        
        &GenAU-S w/ FiT& \textbf{15.76} & \textbf{21.29} & \textbf{0.36} \\

        \bottomrule
        \end{tabular}}
\vspace{-2em}
\end{table}

\textbf{Comparison with baselines.} In \tabref{tab:audio_generation_results}, we report evaluation results. When trained with a \emph{similar size and data scale} to state-of-the-art methods, \audiogenerator achieves superior performance in most metrics, improving IS by $11.1\%$, FAD by $4.7\%$, and $13.5\%$ in $CLAP_{LAION}$. Using the full 10-seconds subset of \audiodataset-XL further enhances the results with over 21.8\% in IS and 20.5\% in CLAP~score. Additionally, to isolate the impact of model architecture from data quality, we conduct a user study in~\tabref{tab:user_study}, where \audiogenerator is consistently preferred over Make-an-Audio-2, even when trained with comparable data quality.

\textbf{How \audiogenerator scale with synthetic data?} To study this, we train \audiogenerator-S (493M params) for 50k steps by fixing AudioCaps and Clotho in the training data and varying the amount of synthetic data from \audiodataset. As reported in~\figref{fig:scaling} (right), increasing synthetic training data consistently improves both IS and FD. Similar improvements are evident in~\tabref{tab:ood_evaluation} where increasing the dataset size significantly boosts all metrics, improving IS by $56.3\%$. User studies in~\tabref{tab:user_study} further support these findings, where a model training with \audiodataset is consistently favored over a model trained only on AudioCaps. Finally, we report in~\tabref{tab:audio_generation_results} the scaling performance on \audiodataset-XL (47M). To ensure a fair comparison, we train on 10-second subset (19.7M) and observe improvements of 12.3\% in FD and 11.6\% in CLAP score compared to training with \audiodataset (800k). 

\textbf{Is caption quality important?} We compare in~\tabref{tab:ood_evaluation} \audiogenerator-S against the same model trainined with WavCaps~\cite{mei2023wavcaps} captions. We observe gains across all metrics, confirming the importance of caption quality and the improvements brought by our high-quality captioner~\audiocaptioner. Interestingly, expanding data size with lower-quality captions offers no significant gains over training on AudioCaps alone, consistent with~\citet{liu2023audioldm2}. 

\textbf{Does \audiogenerator benefit from model size scaling?} Similar to data scaling, increasing model size consistently enhances performance. As shown in ~\figref{fig:scaling} (left), larger models achieve better FD and IS. This is further confirmed in~\tabref{tab:ood_evaluation}, where \audiogenerator-L (1.25B) outperforms \audiogenerator-S (493M) across all metrics with almost 20.5\% increase in IS. User study in~\tabref{tab:user_study} shows a strong preference for \audiogenerator-L over \audiogenerator-S. Unlike previous methods which reported diminishing returns with model scaling~\cite{ezaudio, liu2023audioldm2},~\audiogenerator continues to improve as model size scale.

\textbf{How do FITs compare with other diffusion backbones?} We evaluate the impact of the diffusion backbone by replacing FIT with a UNet~\citep{ronneberger2015unet}, or a DiT~\cite{dit}. In~\tabref{tab:ood_evaluation}, we observe that \audiogenerator with FIT outperforms these alternatives across all metrics. We infer that the FIT architect, with its read and write operations, allocates compute more efficiently to the key segments of the input, making it suitable for ambient audio clips which often include silent or redundant parts.

\newcommand{\tablesize}{\fontsize{9}{11}\selectfont}
\begin{table}[t]
\centering
\caption{User study between various baselines. \% of votes in favor of the baseline to the left.}
\label{tab:user_study}
\resizebox{\linewidth}{!}{
\begin{tabular}{@{}lcccc@{}}
\toprule
Comparison & Realism & Quality & P-Align. & Pref. \\
\midrule
\textbf{Model:} GenAU-L vs GenAU-S & 61.20\% & 58.00\% & 61.20\% & 60.40\% \\
\textbf{Data:} GenAU-L vs GenAU-L (AC) & 60.40\% & 54.80\% & 60.40\% & 59.20\% \\
\textbf{Backbone:} GenAU-S w/o Recap. vs MAD-2 & 64.40\% & 64.00\% & 63.20\% & 64.80\% \\

\bottomrule
\end{tabular}}
\vspace{-2em}
\end{table}

\section{Conclusion}
\label{sec:conclusions}
We take a holistic approach to improve the quality of existing audio generators. Starting by addressing the scarcity of large-scale captioned audio datasets, we propose a scalable and efficient dataset collection pipeline. We then build a state-of-the-art audio captioning method, \audiocaptioner, which leverages audio metadata to annotate a dataset of 47M annotated audio clips tailored for large-scale audio generation. We then built a latent diffusion model based on a scalable transformer architecture which we trained on our re-captioned dataset to obtain \audiogenerator, a high-quality open-source model for audio generation. Our approach opens up possibilities for extending \audiogenerator to other domains, such as speech and music. Additionally, \audiodataset-XL can serve as a joint text-audio-video dataset and broadens novel applications such as text to audio-video joint generation.

\textbf{Limitations and future work.} \audiocaptioner was fine-tuned on AudioCaps, featuring 4,892 unique words, which limits the diversity of our generated captions. Consequently, \audiogenerator may face challenges in accurately generating audio for detailed prompts. Additionally, while \audiodataset is extensive in size, it has only been validated for audio generation. We leave broader analysis on more tasks for future work.

\newpage

\bibliography{main}
\bibliographystyle{icml2025}

\newpage
\appendix
\onecolumn

\section*{Appendix}
\tableofcontents

\section{\audiodataset-XL Details}
\label{sec:autorecap-xl}
This section outlines the collection and filtering processes for \audiodataset-XL.

\subsection{Stage 1: Data Selection}
We selected existing video datasets primarily from YouTube for the ease of accessing automatic transcriptions. Specifically, we chose 73 million videos from the datasets AudioSet~\citep{gemmeke2017audioset}, VGGSound~\citep{chen2020vggsound}, ACAV100M~\citep{lee2021acav100mautomaticcurationlargescale}, VideoCC~\citep{nagrani2022learningaudiovideomodalitiesimage}, YTTEMP1B~\citep{ytyemp1b}, and HDVila-100M \citep{xue2022hdvila}. We select these datasets for their likelihood of containing videos with strong audio-video correspondence.

\subsection{Stage 2: Speech and Music Filtering}
\begin{figure}[htbp]
    \centering
    \begin{minipage}{0.49\textwidth}
        \centering
        \includegraphics[width=\textwidth]{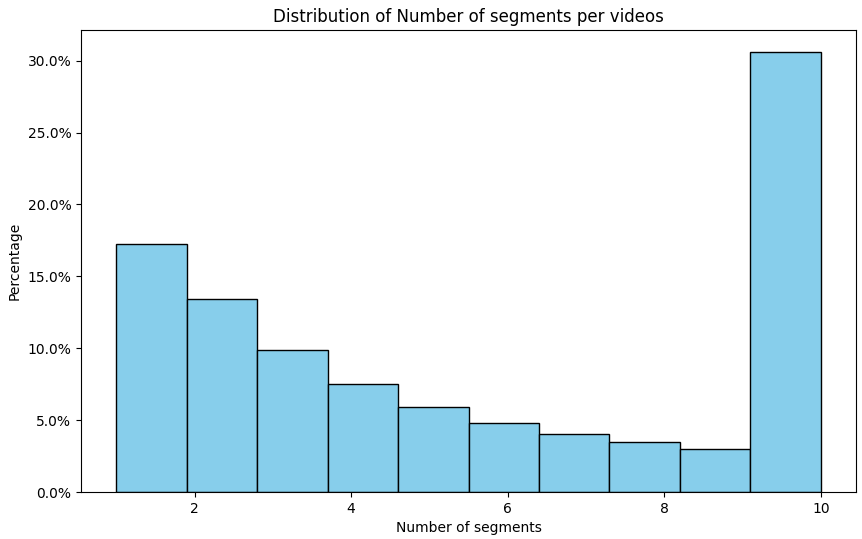} %
        \caption{Distribution of number of segments per-video in \audiodataset-XL-Raw}
        \label{fig:num_segments_pre_filtering}
    \end{minipage}\hfill
    \begin{minipage}{0.49\textwidth}
        \centering
        \includegraphics[width=\textwidth]{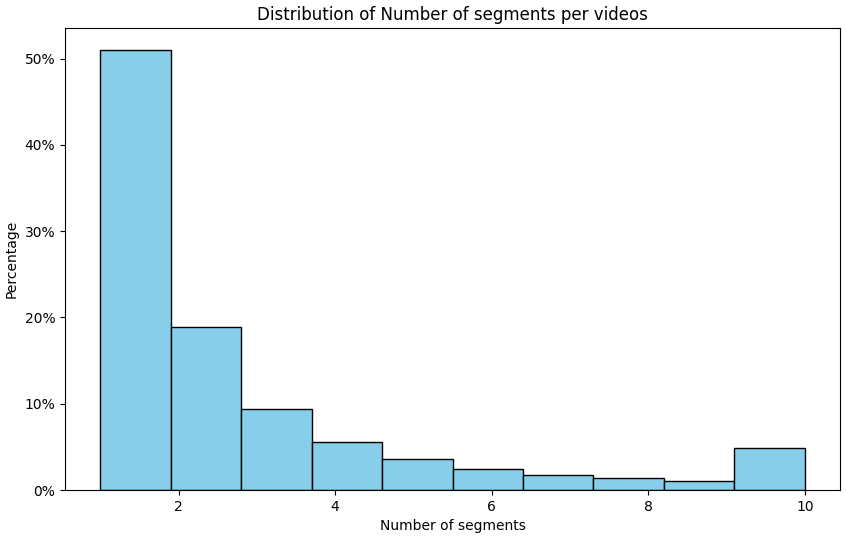} %
        \caption{Distribution of the number of segments per-video in \audiodataset-XL}
        \label{fig:num_segments}
    \end{minipage}
\end{figure}

\begin{figure}[htbp]
    \centering
    \begin{minipage}{0.49\textwidth}
        \centering
        \includegraphics[width=\textwidth]{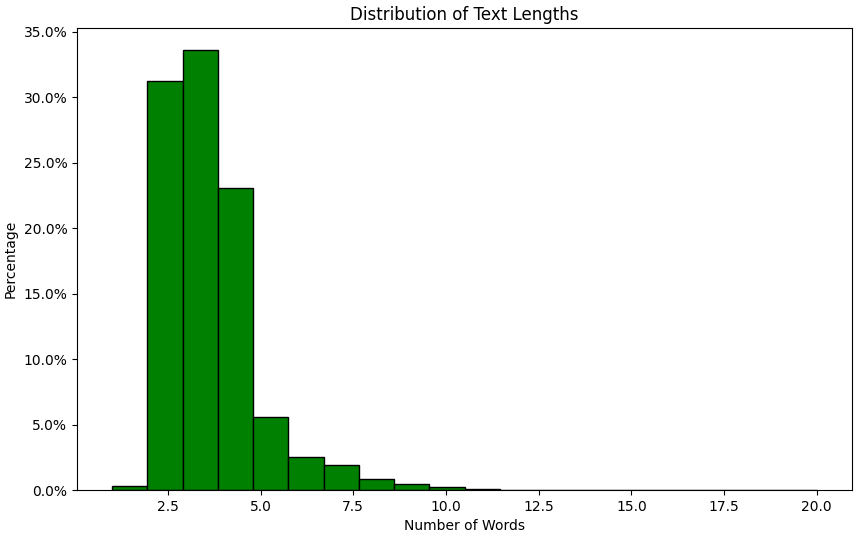} %
        \caption{Distribution of caption length of \audiodataset-XL-Raw}
        \label{fig:caption_length_pre_filtering}
    \end{minipage}\hfill
    \begin{minipage}{0.49\textwidth}
        \centering
        \includegraphics[width=\textwidth]{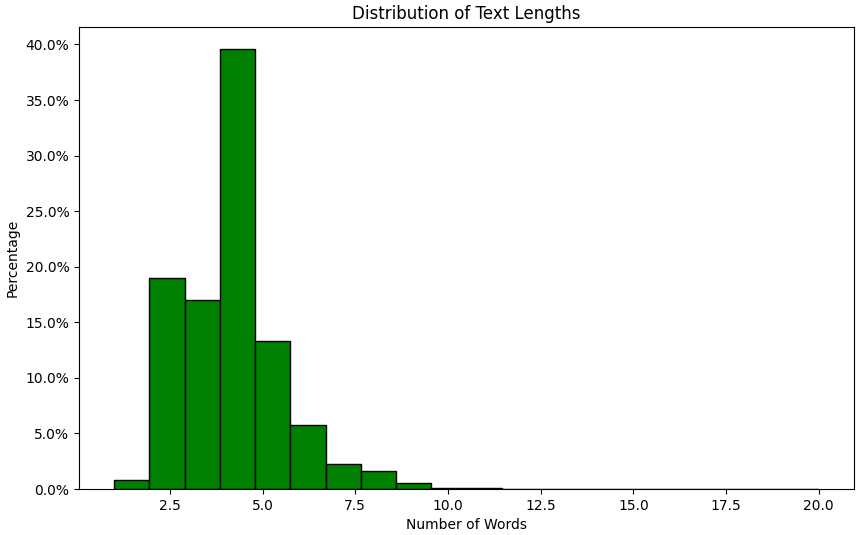} %
        \caption{Distribution of caption length of \audiodataset-XL}
        \label{fig:caption_length}
    \end{minipage}
\end{figure}

\begin{figure}[htbp]
    \centering
    \begin{minipage}{0.49\textwidth}
        \centering
        \includegraphics[width=\textwidth]{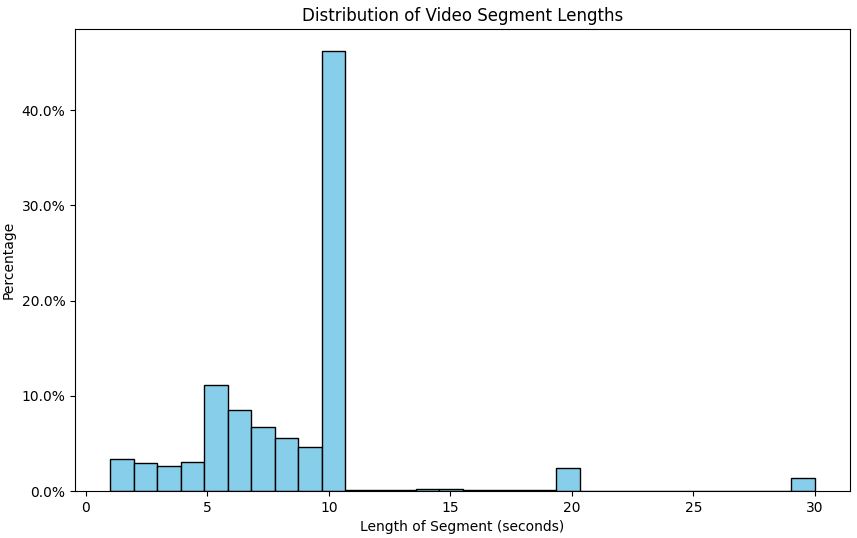} %
        \caption{Distribution of audio duration of \audiodataset-XL-Raw}
        \label{fig:audio_length_pre_filtering}
    \end{minipage}\hfill
    \begin{minipage}{0.49\textwidth}
        \centering
        \includegraphics[width=\textwidth]{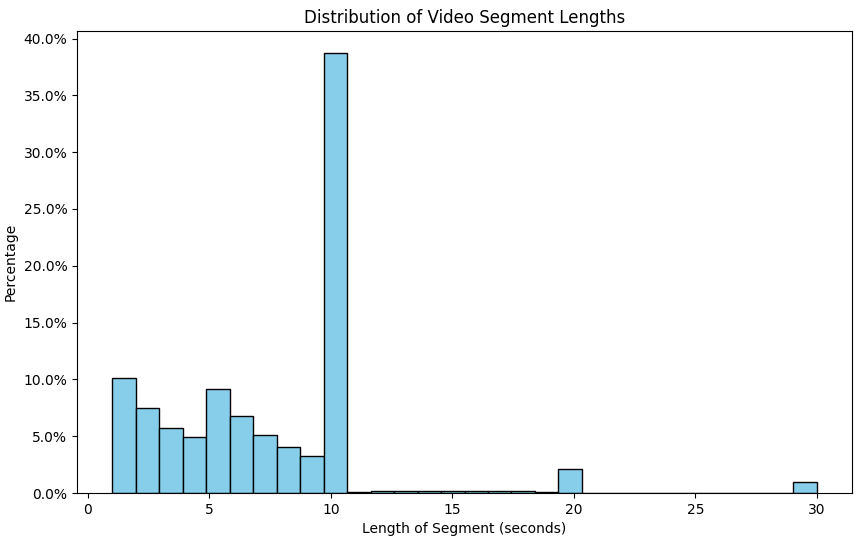} %
        \caption{Distribution of audio duration of \audiodataset-XL}
        \label{fig:audio_length}
    \end{minipage}
\end{figure}

\begin{figure}[htbp]
    \centering
    \begin{minipage}{0.49\textwidth}
        \centering
        \includegraphics[width=\textwidth]{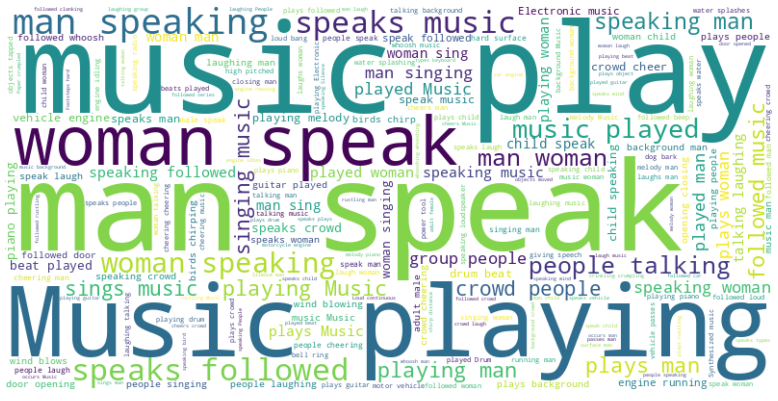} %
        \caption{Word cloud of audio captions in \audiodataset-XL-Raw}
        \label{fig:word_cloud_pre_filtering}
    \end{minipage}\hfill
    \begin{minipage}{0.49\textwidth}
        \centering
        \includegraphics[width=\textwidth]{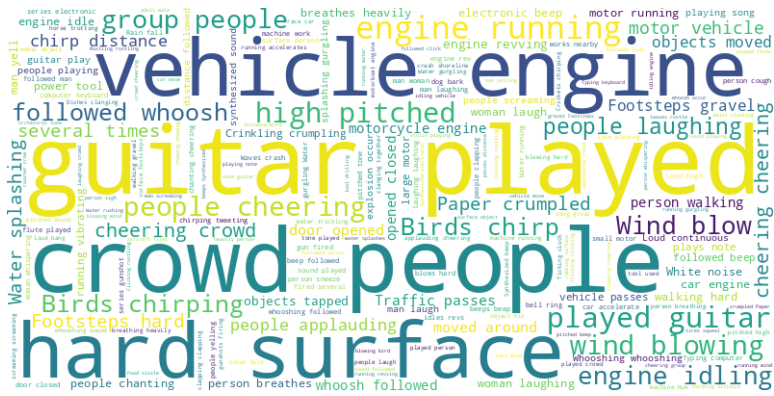} %
        \caption{Word cloud of audio captions in \audiodataset-XL}
        \label{fig:word_cloud}
    \end{minipage}
\end{figure}

We downloaded English transcripts from YouTube and used automatically generated ones for videos without existing transcripts. However, we discard videos without any transcripts. While some datasets provide only video segments with specific timestamps, we processed the full videos, totaling around 73 million videos. We accepted audio segments longer than one second that lacked any corresponding subtitles, indicating the absence of speech and music. After filtering, we isolated approximately 327.3 million segments from 55.1 million videos. \figref{fig:num_segments_pre_filtering} displays the distribution of the number of segments per video. We denote this dataset as \audiodataset-XL-Raw. Subsequently, we use \audiocaptioner to caption the audio segments. \figref{fig:caption_length_pre_filtering} shows the distribution of caption lengths. Given that \audiocaptioner was trained for 10-second audio, we limited segments to this duration. Additionally, we concatenate consecutive segments yielding identical captions to form longer audio clips. \figref{fig:audio_length_pre_filtering} illustrates the audio length distribution, and a word cloud of the captions is shown in \figref{fig:word_cloud_pre_filtering}. Despite filtering, the dataset was still dominated by captions related to speech and music. We attribute this to the limitations of YouTube's automatic transcription, particularly with certain types of music and less common languages.

\begin{figure}
    \centering
    \includegraphics[width=1.0\linewidth]{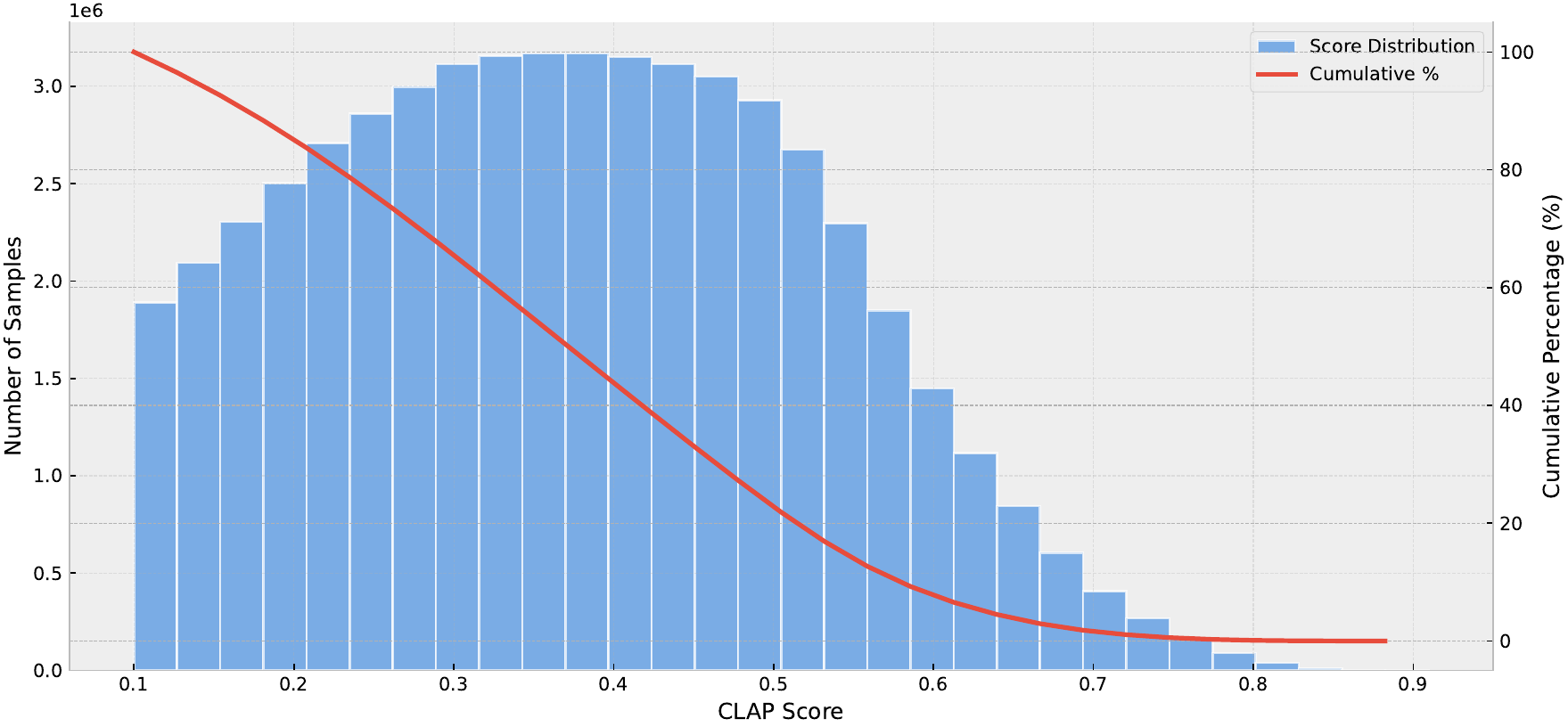}
    \caption{CLAP score distribution of AutoReCap-XL}
    \label{fig:clap-dist}
\end{figure}

\subsection{Stage 3: Post-filtering of Speech and Music.}
To further refine the dataset from speech and music, We follow a simple filtering approach. Specifically, we employed a large language model (LLM) to generate keywords associated with speech and music, such as "talking", "speaking", and "singing," and excluded all audio segments whose captions contained such keywords. Finally, we filter all audio-text pairs with CLAP score less than 0.1. A distribution of the CLAP scores for the collected dataset is present in \figref{fig:clap-dist}.
This process yielded 47 million audio-text pairs from 20.3 million videos. \figref{fig:num_segments} shows the number of segments per video, \figref{fig:caption_length} shows the caption length distribution, \figref{fig:audio_length} shows the audio length distribution, and \figref{fig:word_cloud} presents a word cloud of the final captions. We outline the data sources for constructing this dataset in \tabref{tab:data_details}. Our proposed dataset is not only 75 times larger than the previously largest available dataset, LAION-Audio-630K\cite{wu2023laionaudio} in terms of the number of audio clips, but also provides more accurate captions compared to existing datasets that rely on raw textual data. A comprehensive comparison with other datasets is detailed in \tabref{tab:data_comparision}

\section{Architecture details}
\subsection{HTSAT Embeddings Extraction}
\label{ap:htsat_embeddings_extraction}
\audiocaptioner uses HTSAT~\citep{ke2022htsat} embeddings to encode the input audio and follows the HTSAT-BART~\citep{mei2023wavcaps} embedding extraction procedure, described in the following, to obtain ``fine-grained'' HTSAT embeddings. Given a 10-seconds single-channel input audio at 32Khz, HTSAT represents it as a mel-spectrogram using window size of 1024, 320 hop size, and 64 mel-bins, resulting in an input of shape $(T=1024, F=64)$. The spectrogram is then encoded as latent tokens of shape ($\frac{T}{8P}=32$, $\frac{F}{8P}=2$, $8D=768$) before the classification layer. HTSAT-BART~\cite{mei2023wavcaps}, then averages over the frequency dimension to obtain a representation of shape ($\frac{T}{8P}=32$, $1$, $8D=768$) and replicates the latent token by a token replication factor of $8P = 32$ to obtain a so-called ``fine-grained'' representation of shape $32 \times 32 \times 768$, which is flattened into a representation of shape $1024 \times 768$. We adopt this representation throughout our work, and \apref{ap:additional_htsat_embeddings_evaluation} provides additional evaluation results showing the performance benefits of the token replication operation.

\section{Limitations} 
\subsection{\audiocaptioner} Sounds emitted by various objects can often sound similar, such as a waterfall compared to heavy rain, or a can versus a motorcycle engine. In scenarios where metadata lacks detail, our audio captioning model may struggle to disambiguate these sounds accurately. The model also tends to falter in capturing the temporal relationships between sounds and differentiating foreground from background noises. Additionally, since it is fine-tuned on AudioCaps, which contains a limited vocabulary of 4,892 unique words (excluding common stop words), the model frequently produces repetitive words and captions. 

\subsection{\audiogenerator} Although our model is trained to generate natural sound effects, it underperforms in specialized areas like music generation or text-to-speech synthesis, where more targeted models are superior. Moreover, the limited vocabulary of the paired texts, even though extensive, hampers the model’s ability to accurately generate audio for long and detailed prompts.

\subsection{\audiodataset-XL}
Our proposed dataset, \audiodataset-XL, is substantial in size but features a constrained vocabulary of only 4,461 unique words, excluding stop words, due to the vocabulary limitations of the AudioCaps-trained captioner. Furthermore, despite its potential as a significant contribution, this dataset has not yet been extensively analyzed for caption accuracy or performance in downstream tasks.

\section{Evaluation Details}

\subsection{Audio Captioning}
\label{ap:evaluation_details}
While the established practice in the evaluation of audio captioning methods is to report the results on the test set using the checkpoint that performs best on the validation subset, prior work \citep{labbe2023conette,kim2024enclap} reported high instability of the metrics on the validation subset and weak correlation between the validation and test performance, making the model's results vary significantly for different seeds. To alleviate this, ENCLAP \citep{kim2024enclap} selects around five best-performing validation checkpoints and reports their best results on the test set. CNext-trans \citep{labbe2023conette} uses the FENSE score to pick the best validation checkpoint. This method of choosing the best checkpoint may produce misleading results and potentially disadvantage baselines. Our model, thanks to the two-stage training paradigm, significantly reduces this instability and we observe steady performance gains as training progresses. Therefore, we report the results at convergence, specifically after 20 epochs of pre-training and 20 epochs of fine-tuning. 

\subsection{Audio Generation}

There is a lack of consistency in the metrics used across text-to-audio generation baselines. Some baselines, such as \citet{liu2023audioldm} and \citet{huang2023makeanaudio2}, employ the VGGish network \citep{hershey2017vggish} to compute the Fréchet Distance, while others, like \citet{liu2023audioldm2}, utilize the PANNs network \citep{kong2019pann}, and still others rely on OpenL3 embeddings, such as \citet{evans2024fast}. Additionally, some baselines use the LAION CLAP network \citep{wu2023laionaudio} to compute the CLAP score, whereas others use the Microsoft CLAP network \citep{elizalde2023clap}. To further complicate matters, different baselines often report varying results in various publications. To address these discrepancies, we recalculated all metrics for the baselines using their publicly released checkpoints under identical evaluation configurations. Our method significantly outperforms the baselines across all metrics, except for the Fréchet Distance, where it is slightly behind Make-An-Audio 2 \citep{huang2023makeanaudio2}. Nevertheless, our user study, detailed in the main paper, indicates that \audiogenerator is generally preferred over Make-An-Audio 2.

\begin{figure}
    \centering
    \includegraphics[width=\linewidth]{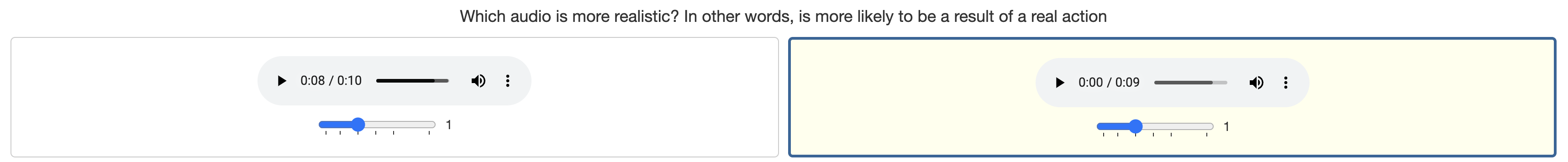}
    \caption{A screenshot of the user study interface.}
    \label{fig:user_study_screenshot}
\end{figure}

\begin{table}[ht]
    \centering
    \begin{tabular}{lr}
        \toprule
        \textbf{Data Source} & \textbf{$\#$ pairs} \\
        \midrule
        AudioSet & 339.4k \\ 
        VGGSounds & 126.9k \\
        Freesounds & 262.3k \\ 
        BBC Sound Effects & 31.2k \\ 
        YouTube Videos & 47.0M \\
        \quad ACAV-100M &  \\ 
        \quad VideoCC &  \\ 
        \quad YTTEMP1B &  \\
        \quad HDVila-100M &  \\
        \midrule
        \audiodataset & 761.1k \\ 
        \audiodataset-XL & 47.0M \\ 
        \bottomrule
    \end{tabular}
    \caption{Overview of the employed dataset sources and audio clips counts for each of them.}
    \label{tab:data_details}
\end{table}
 
\begin{table}[t]
\centering
\renewcommand{\arraystretch}{1.1}
\caption{Comparative overview of the main audio-language datasets.}
\begin{tabular}{lrrc}
\toprule
\textbf{Dataset} &  \textbf{\# Text-Audio Pairs }& \textbf{Duration (h)} & \textbf{Text source} \\ 
\midrule
AudioCaps & 52,904 & 144 & Human \\ 
Clotho & 5,929 & 37 & Human \\
MACS & 3,537 & 10 & Human \\ 
WavText5K & 4,072 & 23 & Online raw-data \\ 
SoundDescs & 32,979 & 1,060 & Online raw-data \\ 
LAION-Audio-630K & 633,526 & 4,325 & Online raw-data \\ 
WavCaps & 403,050 & 7,567 & Processed raw-data \\ \midrule
\audiodataset & 761,113 & 8,763 & Automatic re-captioning \\ 
\audiodataset-XL & 47M & 123,500 & Automatic re-captioning \\
\audiodataset-XL-Raw & 327.3M & - & Automatic re-captioning \\\bottomrule
\end{tabular}
\label{tab:data_comparision}
\end{table}

\begin{table}[ht]
\centering
\caption{Audio Evaluation Criteria}
\label{tab:user_study_sample}
\begin{tabular}{@{}lp{10cm}@{}} %
\toprule
\textbf{Criterion} & \textbf{Description} \\
\midrule
Realism & Which audio is more realistic? In other words, is more likely to be a result of a real action. \\
\midrule
Quality & Which audio has better quality, regardless of the realism of the audio. Please note that some audio may have background noise, which should not be confused with low quality. \\
\midrule
Prompt Alignment & Considering the prompt to generate the audio is "A sewing machine operating as a machine motor hisses loudly in the background", which audio better follows the given prompt? \\
\midrule
Overall Preference & Considering the realism, quality, and prompt alignment of the audio, which audio do you prefer more overall? The prompt is: "A sewing machine operating as a machine motor hisses loudly in the background." \\
\bottomrule
\end{tabular}
\end{table}

\subsection{User Study}
Each user study reported in this paper involved 5 different participants, yielding a total of 1000 responses per study. Samples were selected from the AudioCaps test split, specifically choosing the top 200 samples with the longest text prompts and sampling 50 for each study to enhance the likelihood of obtaining more complex audio scenarios. To minimize discrepancies between baselines, we fix the seed and other sampling parameters across all experiments.

During the user study, participants were initially presented with two audio clips from the compared baselines and asked to judge which one sounded more realistic. They were then prompted to choose the audio they believed had better quality. Next, after showing the prompt used to generate the audio, participants were asked to select the clip that most faithfully followed the prompt. Finally, they were asked to choose their overall preferred audio clip. A screenshot of the user study interface is included in \figref{fig:user_study_screenshot}, and the questions posed to the annotators are detailed in \tabref{tab:user_study_sample}.

\begin{table}[t]
\centering
\small
\setlength\tabcolsep{8pt}
\renewcommand{\arraystretch}{1.1}
\caption{Qualitative comparison of captioning results on the AudioCaps dataset. See the \website~for qualitative results accompanied by the respective audio.}
\label{tab:caption_qualitatives}
\begin{tabular}{@{}lp{115mm}@{}}
\toprule
Method & Caption \\ \midrule
Ground Truth & \emph{A man talking as ocean waves trickle and splash while wind blows into a microphone}   \\
\name &  \emph{A man speaks as wind blows and water splashes} \\
CoNeTTE  &  \emph{A man is speaking and wind is blowing}  \\
ENCLAP & \emph{A man is speaking and wind is blowing}  \\
\midrule
Ground Truth &  \emph{An adult male speaks, birds chirp in the background, and many insects are buzzing}  \\
\name &  \emph{Birds chirp in the distance, followed by a man speaking nearby, after which insects buzz nearby}  \\
CoNeTTE &  \emph{A man speaking with birds chirping in the background.}  \\
ENCLAP  & \emph{Birds are chirping and a man speaks}   \\
\midrule
Ground Truth & \emph{A telephone dialing tone followed by a plastic switch flipping on and off}   \\
\name &  \emph{A telephone dialing followed by a series of plastic clicking then plastic clanking before plastic thumps on a surface}  \\
CoNeTTE & \emph{A telephone ringing followed by a beep.}   \\
ENCLAP  & \emph{A telephone dialing followed by a series of electronic beeps}   \\
\midrule
Ground Truth & \emph{A running train and then a train whistle}   \\
\name & \emph{A train moves getting closer and a horn is triggered}   \\
CoNeTTE & \emph{A train horn blows and a steam whistle is blowing}   \\
ENCLAP & \emph{A train running on railroad tracks followed by a train horn blowing as wind blows into a microphone}   \\
\midrule
Ground Truth & \emph{A child is speaking followed by a door moving}   \\
\name & \emph{A child speaks followed by a loud crash and a scream}   \\
CoNeTTE & \emph{A woman speaking followed by a door opening and closing.}   \\
ENCLAP  & \emph{A young girl speaks followed by a loud bang}   \\
\midrule
Ground Truth & \emph{Water splashing as a baby is laughing and birds chirp in the background}   \\
\name & \emph{A baby laughs and splashes, and an adult female speaks}   \\
CoNeTTE  & \emph{A baby is laughing and people are talking.
}   \\
ENCLAP & \emph{A baby laughs and splashes in water}   \\
\midrule
Ground Truth & \emph{Leaves rustling in the wind with dogs barking and birds chirping}   \\
\name & \emph{Birds chirp in the distance, and then a dog barks nearby}   \\
CoNeTTE  & \emph{A dog is barking and a person is walking.}   \\
ENCLAP & \emph{Birds chirp and a dog barks}   \\
\midrule
Ground Truth & \emph{Tapping followed by water spraying and more tapping}   \\
\name & \emph{Some light rustling followed by a clank then water pouring}   \\
CoNeTTE & \emph{A toilet is flushed and water is running.}   \\
ENCLAP  & \emph{A faucet is turned on and runs}   \\

\bottomrule
\end{tabular}
\vspace{-3mm}
\end{table}

\begin{table}[t]
\centering
\small
\setlength\tabcolsep{2pt}
\renewcommand{\arraystretch}{1.1}
\caption{Ablation of different FIT architectural variations in terms of patch size number of latent tokens and adopted text encoders on the AudioCaps dataset.}
\label{tab:fit_ablation}
\begin{tabular}{@{}lccccccc@{}}
\toprule
Tokens & Patch size & FLAN-T5 & CLAP & FD $\downarrow$  & FAD $\downarrow$ & IS $\uparrow$ \\ \midrule
256 & 1 & \checkmark & \checkmark & \textbf{16.45} & \textbf{1.29} & \textbf{10.26}  \\
\midrule
256 & 1 &  & \checkmark & 17.41 & 1.39 & 10.0  \\
256 & 1 & \checkmark &  & 20.47 & 1.86 & 8.89 \\
\midrule
384 & 1 &  & \checkmark & 17.41 & 1.39 & 10.0  \\
192 & 1 &  & \checkmark & 18.0.1 & 2.01 & 8.91 \\
128 & 1 &  & \checkmark & 25.56 & 1.77 & 7.49 \\
\midrule
256 & 2 & \checkmark & \checkmark & 18.53 & 1.70 & 9.0 \\

\bottomrule
\end{tabular}
\end{table}

\begin{table}[t]
\centering
\small
\setlength\tabcolsep{8pt}
\renewcommand{\arraystretch}{1.1}
\caption{Ablation of different 1D-VAE designs on audio generation on the AudioCaps dataset.}
\label{tab:vae_ablation}
\begin{tabular}{@{}llccc@{}}
\toprule
Channels & Recon. loss & FAD $\downarrow$  & FD $\downarrow$ & IS $\uparrow$ \\ \midrule
64 & 0.159 & \textbf{1.29} & \textbf{16.45} & \textbf{10.26} \\
128 & 0.107 & 1.43 & 16.78 & 10.11  \\
256 & \textbf{0.064} & 1.80 & 18.63 & 9.43 \\
\bottomrule
\vspace{-3mm}
\end{tabular}
\end{table}

\section{Training and Inference Details}

\subsection{\audiocaptioner} 

\audiocaptioner introduces 6.2 million new parameters on top of the frozen HTSAT audio encoder and the base BART model. These parameters include 4.7M for the Q-Former, 0.9M for embedding layers, and 0.6M for projection layers. The Q-Former employs 256 learnable tokens, a hidden dimension of 256, 8 attention heads, and 2 hidden layers.

We train the audio captioning model using the Adam optimizer, starting with a learning rate of $10^{-4}$ in stage 1, and reducing to $10^{-5}$ in stage 2. The training was completed over 9 hours on eight A100 80GB GPUs. Although our model is training with 10-second audio clips, we observed qualitatively that it generalizes well to short audios, such as 1-2 second audio clips. 

\subsection{\audiogenerator} We employ the LAMB optimizer for our audio generation model, setting the learning rate at $0.005$ with a cosine schedule, and incorporating a weight decay of $0.1$ and a dropout rate of $0.1$. The small model variant is trained for 210k steps with a batch size of 2,048, while the large model variant is trained for 220k steps with a batch size of 3,072. The large model is trained over 48 hours on 48 A100 80GB GPUs, and the small model on 32 GPUs. Ablation studies are conducted on eight A100 80GB GPUs using a batch size of 512. We further condition the model on the training dataset with a conditioning dataset ID. For generation, we utilize the AudioCaps dataset ID as it is the most reliable dataset.

\section{Discussion with Concurrent work}
\subsection{Text-conditioned audio generation}
Recently, Stable Audio Open~\citep{stableaudioopen} introduced a 1.32B-parameter model capable of generating variable-length stereo audio clips at 44.1 kHz. This model leverages a latent diffusion approach with a DiT~\citep{dit} as its diffusion backbone, similar to prior work such as Make-An-Audio 2~\citep{huang2023makeanaudio2}. In contrast, \audiogenerator employs a FiT architecture. In~\tabref{tab:user_study}, we show the superiority of our FiT-based approach over DiT by showing that \audiogenerator-S is consistently preferred over a 937M-parameter DiT-based baseline (Make-An-Audio 2~\cite{huang2023makeanaudio2}) when trained on comparable data settings (\ie without recaptioning) at a smaller scale (493M parameters). Additionally, Stable Audio Open proposes directly encoding audio clips using a variational autoencoder (VAE) with a ResNet-like architecture, which is particularly effective for higher-resolution audio generation. In contrast, our work adopts previous approaches~\citep{huang2023makeanaudio2, liu2023audioldm2} and uses a Mel-spectrogram representation due to its simplicity. \audiogenerator, being a latent model, can readily benefit from improved latent audio representations, such as those employed by Stable Audio Open.

\subsection{Audio captioning}
A concurrent work, SOUND-VECAPS~\citep{soundvecapsimprovingaudiogeneration}, and Auto-ACD~\citep{autoacd}, propose prompting a pretrained large language model with multimodal information. SOUND-VECAPS utilizes visual captions generated by a pretrained visual captioner~\citep{cogvlm} alongside audio captions from a pretrained audio captioner, ENCLAP~\citep{kim2024enclap}, to produce more complex captions, showing significant improvements in the downstream task of audio generation. This aligns with our approach of incorporating visual captions in the audio captioning task. However, unlike these methods, which rely solely on pretrained models, we integrate visual information directly into the training process of the audio captioner. This enables a more dynamic and context-aware incorporation of visual information in the audio captioning task.

Additionally, there has been a recent trend toward training large audio-language models~\citep{gama_audio, audioflamingonovelaudio, gong2024listenthinkunderstand, deshmukh2024pengiaudiolanguagemodel} and utilizing them for audio captioning in zero-shot settings. While promising in the pursuit of general-purpose models, their reported results on audio captioning remain inferior to state-of-the-art automatic audio captioning (AAC) methods. Consequently, we opt to train a dedicated AAC model, \audiocaptioner, to achieve the highest-quality captions for our proposed dataset, \audiodataset. 

\section{Additional Results}
In this section, we present additional results which are complemented by our \website.

\subsection{Additional Audio Captioning Evaluation}
\label{ap:additional_captioning_evaluation}

In \tabref{tab:caption_qualitatives} we show qualitative results of the captions produced by our method and compare them with state-of-the-art AAC methods. See the \website~for qualitative results accompanied by the original audio. While ENCLAP \citep{kim2024enclap} and CoNeTTE \citep{labbe2023conette} tend to produce short captions, our method produces the most descriptive captions, capturing the most amount of elements from the ground truth audio, an important capability to allow high-quality audio generation \citep{betkerdalle3}.

\subsection{Additional Audio Generation Evaluation}
\label{ap:additional_generation_evaluation}

In this section, we report additional evaluation results and ablations on the task of audio generation.

In \tabref{tab:fit_ablation}, we evaluate fundamental architectural choices in the design of our scalable FIT model.
When removing either the Flan-T5 or CLAP encodings, we notice a steady reduction in all metrics. When increasing the number of latent tokens we also notice a steady improvement in performance as more compute is allocated to the model. Similarly, increasing the patch size to 2 results in a performance decrease under all metrics due to the reduced amount of allocated computation.

In \tabref{tab:vae_ablation}, we ablate the 1D-VAE bottleneck size in terms of reconstruction loss and performance of a subsequently trained latent audio diffusion model, in terms of FAD, FD, and IS. Similarly to the phenomenon observed in the image and video generation domain \citep{gupta2023photorealistic,esser2024scaling}, we observe that a larger number of channels allocated to the latent space results in lower reconstruction losses, but making the latent space more complex, hindering generation quality. We adopt 64 1D-VAE channels for all our experiments.

\subsection{Additional HTSAT Embedding Extraction Evaluation}
\label{ap:additional_htsat_embeddings_evaluation}

\begin{table}[t]
\centering
\small
\setlength\tabcolsep{8pt}
\renewcommand{\arraystretch}{1.1}
\caption{Ablation of token replication factors for the HTSAT embeddings extraction procedure of~\citep{mei2023wavcaps} on the AudioCaps test split. Larger token replication factors consistently improve performance due to the related compute increase in the downstream model.}
\label{tab:htsat_embeddings_ablation}
\begin{tabular}{@{}lcccccc@{}}
\toprule
& Tokens Count & Replication Factor & CIDEr & BLEU1 & BLEU4 & ROUGE$_L$ \\
\midrule
HTSAT-BART & 32 & 1x & 73.7 & 68.6 & 25.0 & 49.7 \\
HTSAT-BART & 256 & 8x & 74.4 & 69.7 & 26.0 & \textbf{49.8} \\
HTSAT-BART & 1024 & 32x & \textbf{76.6} & \textbf{71.5} & \textbf{26.3} & \textbf{49.8} \\
\midrule

\audiocaptioner{} & 32 & 1x & 81.9 & 71.7 & 28.9 & 51.3 \\
\audiocaptioner{} & 1024 & 32x & \textbf{82.7} & \textbf{72.5} & \textbf{29.3} & \textbf{52.0} \\

\bottomrule
\end{tabular}
\end{table}

\begin{figure}[ht]
    \centering
    \begin{tabular}{cc}
        \includegraphics[width=0.9\linewidth,trim={0 7mm 0 0},clip]{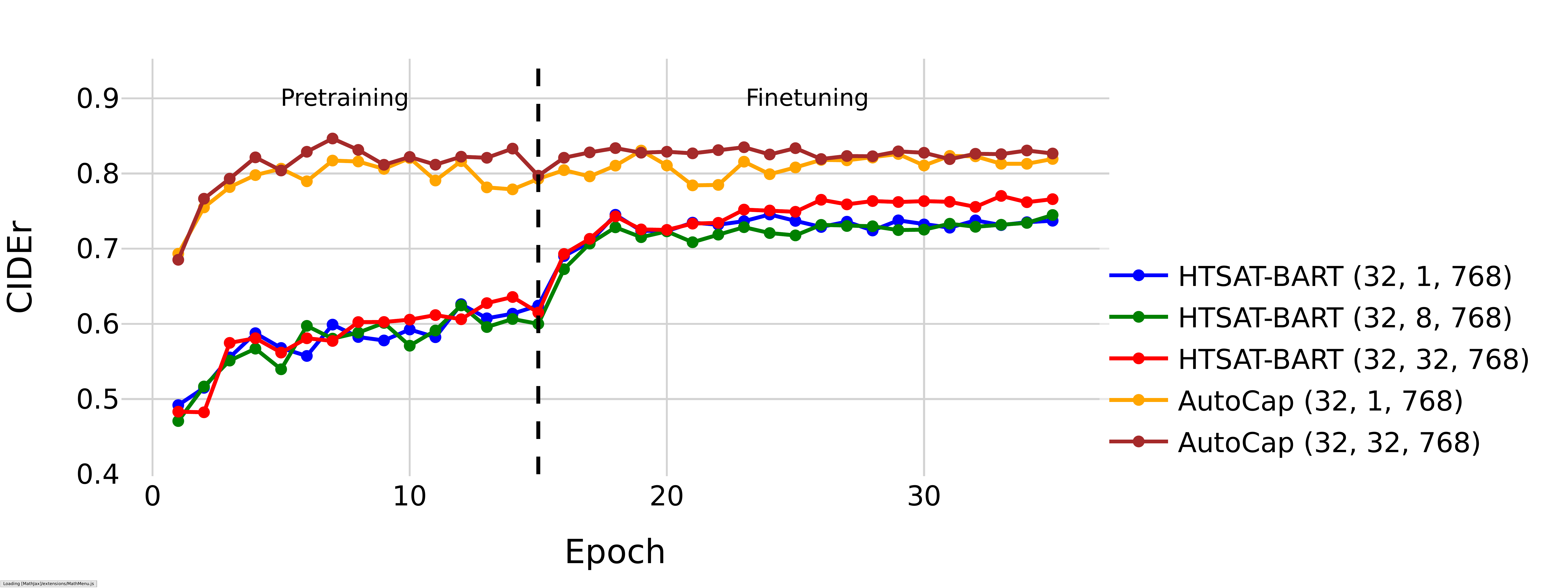} \\ 
        \includegraphics[width=0.9\linewidth,trim={0 7mm 0 0},clip]{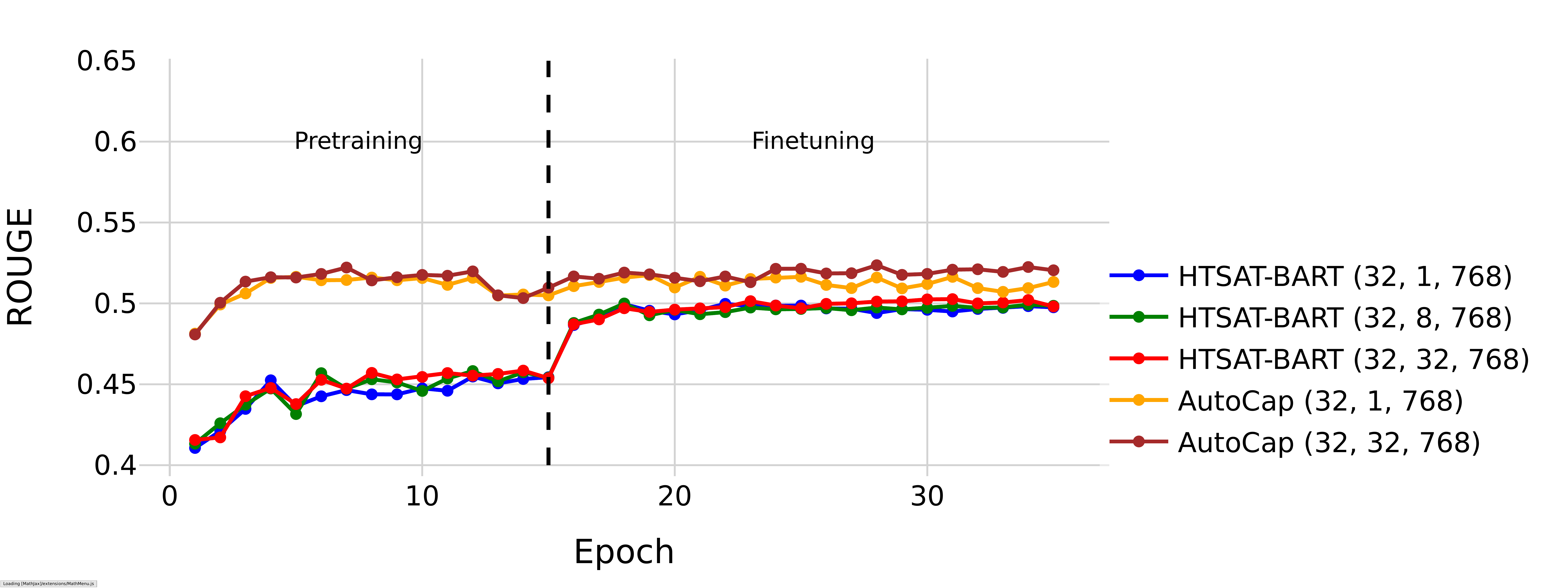} \\
        \includegraphics[width=0.9\linewidth,trim={0 7mm 0 0},clip]{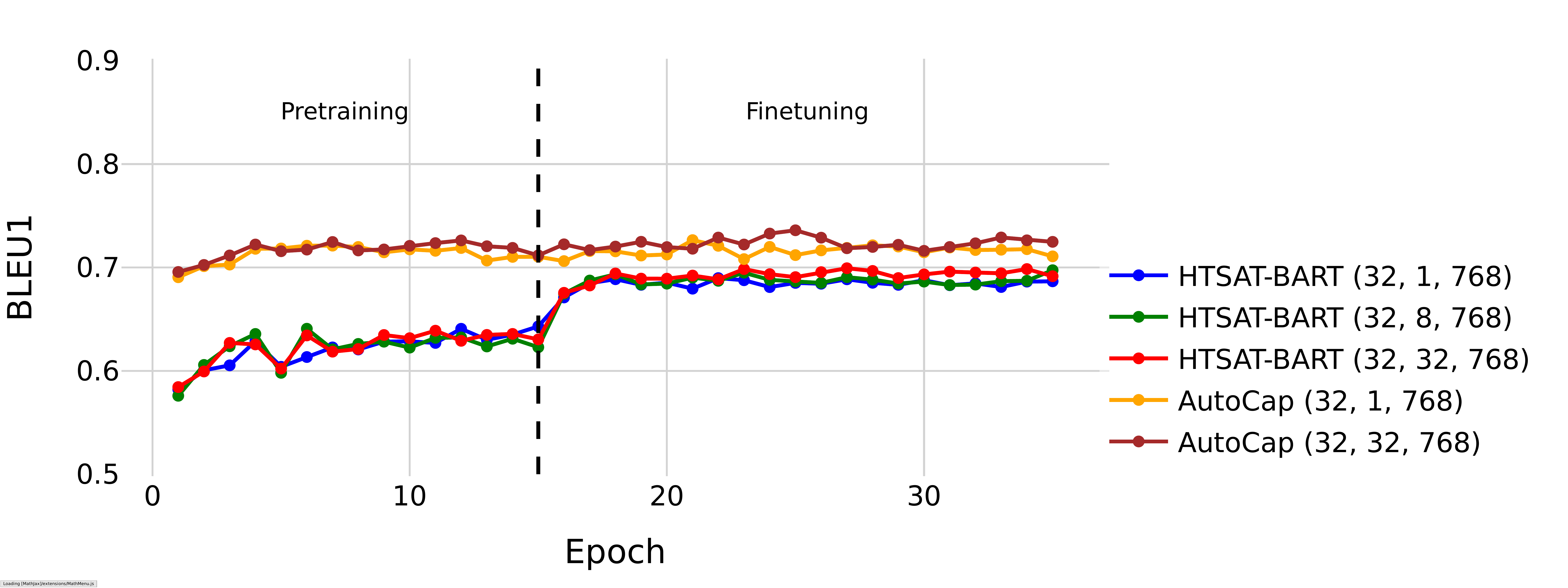} \\ 
        \includegraphics[width=0.9\linewidth,trim={0 7mm 0 0},clip]{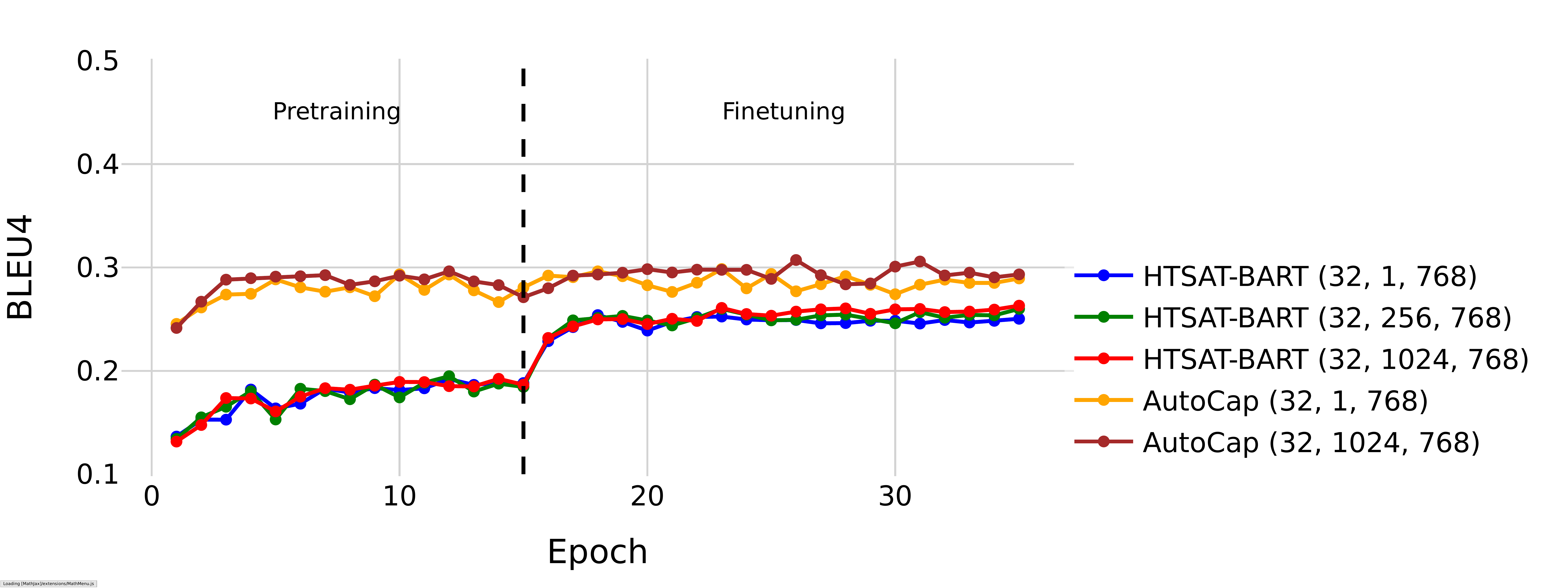} \\
    \end{tabular}
    \caption{Ablation of token replication factors for the HTSAT embeddings extraction procedure of~\citep{mei2023wavcaps} on the AudioCaps test split for the HTSAT-BART~\citep{mei2023wavcaps} and our \audiocaptioner{} model.}
    \label{fig:htsat_embeddings_ablation_combined}
\end{figure}

We perform a series of ablations on HTSAT-BART~\cite{mei2023wavcaps} employing different variants of the procedure of~\cite{mei2023wavcaps} for the extraction of HTSAT embeddings (see~\apref{ap:htsat_embeddings_extraction}). We consider HTSAT output tokens of shape $32\times768$ after the averaging operation over the frequency dimension of~\cite{mei2023wavcaps}, and apply different token repetition factors to produce embeddings with 32 tokens (no token repetition), 256 tokens (8x token repetition) and 1024 tokens (32x token repetition following~\cite{mei2023wavcaps}). For completeness, we perform the same ablation on our \audiocaptioner{}, using as input to the Q-Former 32 tokens (no token repetition) and 1024 tokens (32x token repetition). Training hyperparameters of \audiocaptioner{} are modified to match HTSAT-BART~\cite{mei2023wavcaps} for the purpose of the ablation.

We followed the training procedure of \cite{mei2023wavcaps} and report evaluation results on the AudioCaps test split for the last obtained checkpoint in \tabref{tab:htsat_embeddings_ablation} and \figref{fig:htsat_embeddings_ablation_combined}. As the ablation shows, the token replication operation consistently improves model performance. We attribute this finding to the increased computation in the downstream model caused by it and consequently adopt the best performing 32x token replication embeddings extraction procedure of \cite{mei2023wavcaps} throughout our work.

\end{document}